\documentclass[usenatbib]{mn2e}
\usepackage{graphicx, amsmath, amssymb,verbatim}

\newcommand{\mnred}{M_{N_{\rm red}}}
\newcommand{\mnph}{M_{N_{\rm ph}}}
\newcommand{\msz}{M_{\rm SZ}}
\newcommand{\mwl}{M_{\rm WL}}
\newcommand{\mvel}{M_{\rm Vel}}
\newcommand{\mtrue}{M_{\rm true}}

\begin{document}
\topmargin-1cm

\title{Disentangling correlated scatter in cluster mass measurements}

\author[Noh and Cohn]{Yookyung Noh${}^{1}$ and J.D. Cohn${}^2$\\
${}^1$ Department of Astronomy and Theoretical Astrophysics Center, University of California, Berkeley, CA 94720\\
 ${}^2$ Space Sciences Laboratory and Theoretical Astrophysics Center, University of California, Berkeley, CA 94720}

\date{\today}

\maketitle

\begin{abstract}
The challenge of obtaining galaxy cluster masses is increasingly being
addressed by multiwavelength measurements.
As scatters in
measured
cluster masses are often sourced by properties of or around the
clusters themselves, correlations between mass scatters are frequent
and can be significant, with consequences for errors on mass estimates
obtained both directly and via stacking.  Using a
high resolution $250 h^{-1} Mpc$ side N-body simulation, 
combined with proxies for observational cluster
mass measurements, we obtain mass scatter
correlations and covariances for 243 individual clusters along $\sim
96$ lines of sight each, both separately and together.  Many of these
scatters are quite large and highly correlated.  We use
principal component analysis (PCA)
to characterize scatter trends and variations between clusters. 
PCA identifies combinations of scatters, or variations more generally,
which are uncorrelated or non-covariant.  The PCA combination of mass measurement
techniques which dominates the mass scatter is similar for many
clusters, and this combination is often present in a large amount when
viewing the
cluster along its long axis.
We also correlate cluster
mass scatter, environmental and 
intrinsic properties, and use PCA to find shared trends between
these.  For example, if the average measured richness, velocity
dispersion and Compton decrement mass for a cluster along
many lines of sight are high relative to its true mass, in our simulation
the cluster's mass measurement scatters around this average are also high, its
sphericity is high, and its triaxiality is low.

Our analysis is based upon estimated mass distributions for fixed
true mass.  Extensions to observational data would require further
calibration from numerical simulations, tuned to specific observational
survey selection functions and systematics.

\end{abstract}

\section{Introduction}
Although there is no question that galaxy clusters are the most
massive virialized objects in the Universe, identifying the mass of
any particular cluster remains a challenge.   A cluster's
mass is however one of its core properties, important for using cluster
samples statistically to constrain cosmological parameters, for
understanding clusters as hosts for galaxy evolution, and for studying
the growth and other properties of the clusters themselves.
Recent reviews include \citet{Voi05,Bor06,Mye09,AllEvrMan11}.

In simulations, the cluster mass is the sum of the masses of
the simulation particles which are cluster members, for whatever
cluster
member (and thus mass) definition is used (comparisons of some
mass definitions are found in, e.g., \citet{Whi01,TreePM,HuKra03,CuePra08,Luk09}).
From a given cosmological set of parameters, simulations predict well
defined and 
directly measurable masses, accurate to the
extent that the simulation captures the required physics and has the
requisite resolution. 
These theoretical mass definitions cannot be directly applied to
observational data, as observations instead measure
properties of galaxies both within the cluster and near the cluster,
or gas within and sometimes near the cluster, 
or the bending of space by mass in and around the cluster.  These
observational
mass proxies are converted, via physical modeling and assumptions, to
mass measurements for comparison with theory.
Improvements in the noisy mappings between observationally accessible cluster
properties and theoretically calculable cluster 
properties is much sought after by both theorists and observers.
Complications for observations include projection (and more
generally the lack of three dimensional information\footnote{There is
a long history, e.g. for
optical cluster richness
starting with \citet{Abe58} and continuing with, for example,
\citet{Dal92,Lum92, vHaarlem97,Whi99,CohEvrWhi07,CohWhi09,Roz11,Bie12}, for cluster
weak lensing (e.g.,
\citet{RebBar99,MetWhiLok01,Hoe01,PutWhi05,Men10,BecKra11,Hoe11}), for 
cluster 
Sunyaev-Zel'dovich \citep{SunZel72, SunZel80} (SZ) flux measurements,
(e.g., \citet{WhiHerSpr02,HolMcCBab07, Hal07,ShaHolBod08,Ang12}) and for
cluster velocity dispersions (e.g.,
\citet{Cen97,Tor97,KasEvr05,Biv06,WCS,Sar12}).}) and reliance of
the mapping between mass and observational proxy upon simplifying assumptions such
as hydrostatic equilibrium.  The simulation based theoretical
approaches, for their part,
find it challenging to capture the directly
observable baryonic physics, including galaxy properties.

In order to alleviate systematics and reduce errors in observationally
obtained cluster masses, it is becoming common to combine
measurements from different (often multiwavelength) observational techniques.  The
advantages of complementary information and crosschecks are unfortunately
mitigated by the fact that scatters from different observational
methods are often correlated.  Essentially, as physical properties of the clusters
themselves and their environments are often the causes of mass
measurement scatters, more than one measurement technique can
be affected.  It is important include these correlations in order to
properly estimate the errors in mass measurements of any individual cluster and  
to avoid a bias when stacking clusters on one property and
measuring another. (For discussion see
\citet{Ryk08,CohWhi09,Sta10,WCS}, additional simulated examples of correlated
scatters using different
observational methods include \citet{Men10,Ras12}; analyses are
beginning to include these, e.g. \citet{Roz09,Man10,Ben11}.).  A recent
application to an observational cluster sample, resolving some
questions raised by earlier analyses is found in \citet{Ang12}.)

Here we consider multiwavelength mass measurements for clusters ``observed'' in a cosmological dark
matter simulation.   Our primary focus is on mass scatters for
 individual clusters viewed along several different lines
of sight.  We measure and characterize the multiwavelength correlations and
covariances, and study their relation to other cluster properties using
both correlations and PCA, principal component analysis.
This extends recent work using PCA
to compare relationships between cluster \citep{SkiMac11,Jee11}
properties such as concentration, mass and
ellipticity in simulations, and some supercluster counterparts, \citet{Ein11,Ein12}, in observations.\footnote{Comparing cluster mass scatters to
physical cluster properties has a long history, recent studies include
\citet{YanBhaRic10,BecKra11,BatBonPfrSie11, BahMcCKin12}, as well as papers
mentioned above.} 

The mass observables we simulate are
red galaxy richness, phase space richness, velocity dispersions,
Sunyaev-Zel'dovich decrement and weak lensing $\zeta$ statistic \citep{Fah94,Kai95}, techniques in use or planned for large volume
current and upcoming cluster surveys such as Atacama Cosmology Telescope (ACT
\footnote{www.physics.princeton.edu/act/}), South Pole Telescope
(SPT\footnote{pole.chicago.edu}), Blanco Cosmology Survey
(BCS\footnote{cosmology.uiuc.edu/BCS/}), Dark Energy Survey
(DES\footnote{www.darkenergysurvey.org}) and
Large Synoptic Survey Telescope (LSST\footnote{www.lsst.org/}).

The mock simulation measurements, their scatters, and general
information about
PCA are in \S 2.
(Much of \S 2 summarizes work on the same
simulation detailed in \citet{WCS} (hereafter WCS), and further studied in \citet{NohCoh11} and 
\citet{Coh11}.)
In \S 3, mass scatters for each cluster,
along $\sim 96$ lines of sight, are correlated, and their covariances
are analyzed via PCA.  Distributions of the scatter properties are
considered, and, cluster by cluster,  the PCA direction of largest scatters is
compared to special physical cluster directions.
In \S 4 cluster properties, including the individual cluster mass scatter
distributions, and
environmental and intrinsic properties, are intercompared using correlations and PCA.
In \S 5 PCA is instead applied to mass scatter for the whole sample of
clusters at once, to analyze scatter including both line of sight and
cluster-to-cluster variation, with some discussion of possible
extensions to observations.
\S 6 discusses outliers and \S 7  summarizes.

While we were preparing this work for publication, \citet{Ang12}
appeared.  They consider correlated mass scatter in multiwavelength
measurements, in a 4.1 Gpc side simulation which also includes X-ray.

\section{Simulations and Methods}
\subsection{N-body data}
Our simulation data are the outputs of 
an N-body simulation of M. White, described in
detail in WCS.  His TreePM \citep{TreePM} code was run 
with $2048^3$ particles 
in a periodic box with side length $250$ $h^{-1} Mpc$.  The 
45 outputs are equally spaced in $\ln(a)$ from $z=10$ to $z=0$.
Cosmological parameters were taken to be
$(h,n,\Omega_m,\sigma_8)=(0.7,0.95,0.274,0.8)$, consistent with a large
number of cosmological observations.  We focus here on mock observations 
at $z=0.1$, where our methods have
been most closely tuned to and tested with observational data, as
reported in WCS.
Halos are found via Friends of Friends (FoF) \citep{DEFW}, with
linking length $b=0.168$ times
the mean interparticle spacing (connecting regions with density 
at least roughly 100 times the mean background density).  Clusters are halos with FoF masses $M\geq 10^{14} h^{-1}
M_\odot$ ($M$ hereon will mean this $b=0.168$ FoF mass, we will also write this as
$\mtrue$ when comparing to estimates).  There are 243 clusters in the
box.  Note that because we have a periodic box we do not need to
worry about clusters located near the edge, similarly, because we are
using FoF as a halo finder, every particle is uniquely assigned to a
single
halo.

Galaxies are taken to be resolved subhalos, which are
found via Fof6d \citep{DieKuhMad06}, with the implementation as described in
the appendix of WCS.  Subhalos are tracked (see
\citet{WetCohWhi09,WetWhi10} for particular details) from
their infall into their host halos in order to assign luminosities via subhalo
abundance matching \citep{ConWecKra06}.  The resulting galaxy
catalogue minimum luminosity at $z=0.1$ is
$0.2L_*$ (again see WCS for more discussion and validation tests of the
catalogue galaxy properties with observations).
\subsection{Cluster mass measurements and scatters}
\label{sec:measscat}
We consider five cluster mass measurement methods with this simulation (see
WCS for specifics):
\begin{itemize}
\item $N_{\rm red}$:  Richness using the \citet{maxBCG} MaxBCG algorithm based
upon colors.  Galaxy colors are assigned
using the algorithm of \citet{SkiShe09} with evolution of \citet{SP1,SP2,SP3}.
Galaxies are taken to be ``red'' if they have $g-r$ within 0.05 of the peak of the red
galaxy $g-r$ distribution specified by \citet{SkiShe09} for their
observed $M_r$, 
again see WCS for more detail.
\item $N_{\rm ph}$: Richness based upon spectroscopy, with  cluster
  membership assigned via the criteria of \citet{YanMovan07}.
\item SZ:  SZ flux (Compton decrement) is assigned to every
  particle by giving it a temperature
based upon the mass of its halo.
For every cluster, its measured
SZ flux is then the
flux within an annulus of radius $r_{180b}$ (the radius within which the
average mass is greater than or equal to 180 times background
density), through the length of the box, apodized at the edges.  This
was shown in e.g. \citet{WhiHerSpr02} to well approximate hydrodynamic
simulation results for SZ at the scales appropriate for two cluster
surveys mentioned earlier, SPT and ACT.
\item Vel: Velocity dispersions calculated via the method detailed in
  WCS, and based on 
\citet{denHartog, Biv06, Woj07}.
Phase space information is used
to reject outliers and the mass estimate includes the harmonic radius (calculated as
part of the outlier rejection, more details and definitions in WCS).
\item WL: Weak lensing using a singular isothermal sphere (SIS) or NFW model to assume a cluster lens
  profile and then fitting the projected mass, using the $\zeta$ statistic \citep{Fah94,Kai95}, in a cylinder with
radius $r_{180b}$ and (apodized) length of the box (again WCS
describes fitting models, etc.).
\end{itemize}

The red galaxy richness, phase space
richness and velocity dispersions measured in our simulations are
expected to include 
the majority of
systematics that are present in real observations.  The weak lensing
and Compton decrement (SZ) observations however do not include all known
systematics, such as miscentering, shape measurement and source redshift errors for
lensing, and foreground and point source
removal for Compton decrement.
The relatively small box size (250 $h^{-1} Mpc$ on a
side) also means that line of sight scatter is underestimated
(e.g., \citet{WhiHerSpr02,Hal07,HolMcCBab07,CohWhi09,Ang12} for SZ and
\citet{RebBar99,MetWhiLok01,Hoe01,PutWhi05,Men10,BecKra11,BahMcCKin12,Hoe11}
for lensing).

We extend the cluster sample used in WCS 
to a lower mass range , $M\geq 10^{14} h^{-1} M_\odot$, as in 
\citet{NohCoh11,Coh11}.    The five observables listed above are found along 96 lines of
sight for each cluster, each time placing the cluster at the center of the
periodic box.  Just as in WCS,
lines of sight for clusters are removed for all measurement methods when
a more massive cluster has its center within $r_{180b}$ along the line
of sight (this removes $\sim 400$ of the original $\sim$23000 lines of sight).
In addition, to allow fair intercomparisons, only lines of sight
which have reliable mass measurements for all methods are
included;
the $\sim 90$ lines of sight with fewer than 8 galaxies making the cut for a velocity
dispersion estimate, or
either richness $<1.1$ are also removed.  These cuts will have some
effect on the scatters we consider but would be expected to be
identifiable
observationally.

We take the logarithm of these
observables and that of the true mass $M$ to
find the mean relations for all clusters with $M \geq 10^{14}
h^{-1} M_\odot$.  We find relations for bins of $M$ vs. observables, because of the
large scatter at low mass.  The fits are done by throwing out 3 $\sigma$ outliers for three iterations.   
This gives us our map between the observables and 
mass estimates $\mnred,\mnph, \msz,\mvel,\mwl$.

\begin{figure}
\begin{center}
\resizebox{3.5in}{!}{\includegraphics{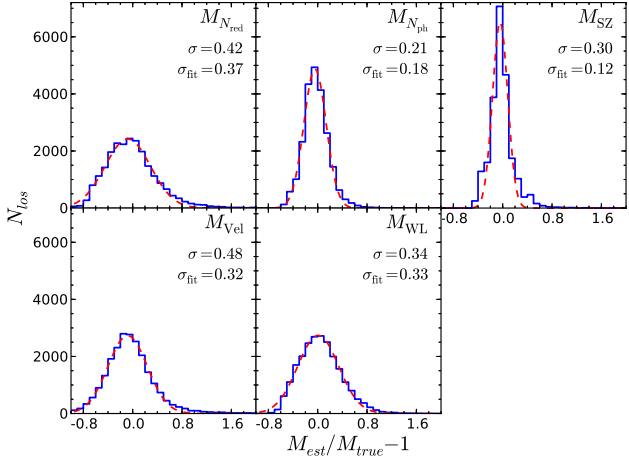}}
\end{center} 
\caption{The fractional mass scatter, $M_{\rm est,i}/\mtrue-1$ for
all 243 clusters along $\sim$96 lines of sight each, for the five mass
measurement methods we consider.  Solid lines are the masses
estimated via $N_{\rm red}$,$N_{\rm
  ph}$,
SZ (Compton decrement), velocity dispersion and WL (weak lensing) as
described
in section \ref{sec:measscat}.  
An approximately 0.2-0.5 standard deviation is
found.  The (for the most part difficult to
distinguish) dashed line is a least squares Gaussian fit, giving
$\sigma_{\rm fit}$, which is much smaller than the standard deviation
for SZ. 
The WL and SZ mass scatter is expected
to be strongly underestimated compared to true observations, 
as the 250 Mpc/h box size is too small to
include all contributions to mass scatter along the line of sight to
the observer, and in these two cases several known systematics are not
included as well.  See text for more information.  
}
\label{fig:fullscatter}
\end{figure}

The distribution of
the fractional mass scatters, $(M_{\rm est}-M_{\rm true})/M_{\rm true}$,
for the five mass measurement methods along $\sim 96$
lines of sight for each of the 243 clusters, is shown in
Fig.~\ref{fig:fullscatter}.
The scatters range from $\sim$ 0.2-0.5, with the smallest mass scatters associated with
Compton decrement\footnote{Recall the caveats for both SZ and 
  weak lensing measurements mentioned earlier, however.} and $N_{\rm
  ph}$.
Note that this work follows a theoretical approach where all the
mass measurements are related to the known true mass, a quantity
inaccessible
in observations.  In particular, the sample is cut on the unobservable
value
of $M_{\rm true}$ and the distributions we consider are
$M_{\rm est}(\mtrue)$, not vice
versa.\footnote{We thank E. Rozo for extremely helpful discussions on this.}  
For an observational sample based upon some measurement $M_{\rm est}$,
such as richness mass or weak lensing mass, often the quantity of
interest is the scatter in $P(\mtrue|M_{\rm est})$ (see also extended
comments in
\S \ref{sec:ensemble}).\footnote{For example, the richness mass scatter
in Fig.~\ref{fig:fullscatter}, looks essentially Gaussian, unlike the double peaked
distribution found and studied in
e.g. \citet{CohEvrWhi07,EriCunEvr11}. 
If looking at $P(\mtrue|M_{\rm est})$, one will
find clusters which are ``blends'', i.e. several halos which
contribute to one apparent halo, often with a higher $M_{\rm est}$
than any of the contributing halos.  These blends (see
also e.g. \citet{Ger05} for discussion) 
are the source
of the bimodal mass distribution reported and used in these other
papers.}

\subsection{Filaments and Galaxy Subgroups}
\label{sec:filgroupdef}
In section \S \ref{sec:scalarprops}, cluster filament properties
calculated in \citet{NohCoh11} and cluster galaxy subgroup
properties calculated in \citet{Coh11} for this simulation are used.  Detailed background
can be found in those two papers, but we briefly summarize some key
aspects here.

Filaments are found in \citet{NohCoh11} using a modification of the
dark matter halo based filament finder of
\citet{Zha09}.\footnote{Although 
the cosmic web was noted years ago
\citep{ZelEinSha82, ShaZel83, Ein84, BonKofPog96}, no unique filament finder exists.  A variety of
finders are in use,  based on a wide range of dark
matter, halo and/or galaxy properties, including for example
\citet{BarBhaSon85, MecBucWag94, SahSatSha98, Sch99, ColPogSou00,
  She03, ColKruCon05,Pim05a, Pim05b, Sto05, NovColDor06, Ara07a, Col07,
  Hah07,vanSch07, Sou08b, StoMarSaa08,BonStrCen09, For09, GonPad09, Pog09,
  SouColPic09, StoMarSaa09, Wu09,Zha09,AraShaSza10,AravanJon10,BonStrCen10,
  MurEkeFre10, Sou10,WayGazSca10,Sha11,ShaHabHei11,Gen12,Smi12,JasWan12}. }
This filament finder searches for bridges 10 $h^{-1} Mpc$ or smaller between halos above $3 \times
10^{10} h^{-1} M_\odot$, starting with the most massive halos as
potential bridge endpoints. 
Some clusters (16/243) end up within filaments because of the finder,
such as
less massive clusters located between two close ($< 10 h^{-1}Mpc$)
massive clusters and  clusters closer than $3 h^{-1} Mpc$ to a larger
cluster.  The rest of the clusters
each lie at the center of a filament map extending out to 10 $h^{-1} Mpc$.
We found that filaments, halo mass (halos with mass  
$\geq 3 \times 10^{10} h^{-1} M_\odot$) and
galaxy richness all tended to lie in a planar region around each
cluster.  We characterized these regions by taking a fiducial
3 $h^{-1} Mpc$ high disk centered on the cluster which extends
out to the edge of the radius 10 $h^{-1} Mpc$ sphere.  We use the
planes related to halo and filament mass below.  For halo mass,
we randomly sampled
10,000 orientations to maximize the halo mass fraction in the plane
(relative to the halo mass in the 10 $h^{-1} Mpc$ sphere).  For
filaments, we considered planes spanned by the cluster and pairs of
filament endpoints, and then took the plane which enclosed the most
filament mass.  This plane was not found for clusters lying within filaments.  See
\citet{NohCoh11} for more details.
We consider four quantities from this analysis below: 
$f_{M_{hplane}}$ (halo mass fraction
in plane relative to that in 10 $h^{-1}M_\odot$ sphere), 
$f_{M_{fplane}}$ (cluster filament mass fraction in
plane relative to sphere), and the respective plane
normal
directions $\hat{n}_{\rm mass},\hat{n}_{\rm fil}$.

Galaxy subgroups were characterized in \citet{Coh11} for this simulation.
These are groups of galaxies that fell into a cluster as part of a shared
halo at an earlier time.  Within the clusters, they share some coherence in space
and
time which can remain for several Gyr.  
We will use for each cluster its largest (richest) galaxy 
subgroup, in particular its fractional richness relative to that of the cluster,
$f_{R_{\rm sub}}$, its displacement relative to the cluster center,
divided by the cluster long axis $f_{D_{\rm Sub}}$, and the directions
of its position and average velocity relative to the cluster
center, $\hat{r}_{\rm sub}, \hat{v}_{\rm sub}$.

\subsection{Principal Component Analysis}
\label{sec:pcadef}
For context and background, we summarize PCA and our notation here
(see, e.g. \citet{Jol02} for extensive discussion).
PCA can be used when there are several correlated or covariant quantities.
It is essentially
a rotation of axes to find linearly independent bases (i.e. quantities
which are not covariant or correlated), and is
based on a model where some underlying
average linear relation is present. 
We will apply PCA in a few different contexts.

Our starting application will be for individual clusters.
For each individual cluster and line of sight,
we have several different methods to estimate the true cluster mass.
Each line of sight can thus be associated with five numbers, where each
number
is the mass
measured in one method.  These numbers can then be thought of as
coordinates in some five dimensional abstract space, with each axis in
this space corresponding to a different measurement method.  
All of the different lines of sight considered together then give a
cloud of points in this space of measurement methods.
PCA gives the properties of the ``shape'' of the mass scatters in this
space, around their
average values for the combined observations,
for each cluster.  We will consider these shapes and how they relate to other cluster properties.
In addition, correspondences can be found between large mass scatters and
physical properties or directions of the cluster.  The PCA direction
with smallest mass scatter is useful as well.
Taking the ensemble of clusters, groups of
properties which change together can be inferred by using PCA on the
full set of correlations (this latter approach was pioneered by \citet{Jee11,SkiMac11,Ein11}).  

There are other uses of PCA, and caveats as well.
PCA is often used to find the minimum set of variables needed to describe a
system to some accuracy, for instance the dominant contributing basis vectors
composing a galaxy spectrum.   As far as caveats go, one concern is that
if correlated variables
are not scattered around a linear relation, a simple rotation of basis
using PCA will not
usefully separate them.
For this reason, sometimes other functions are used besides the
variables themselves, e.g. logarithms, when it is suspected that
variables might be related by power laws.

\begin{figure}
\begin{center}
\resizebox{3.5in}{!}{\includegraphics{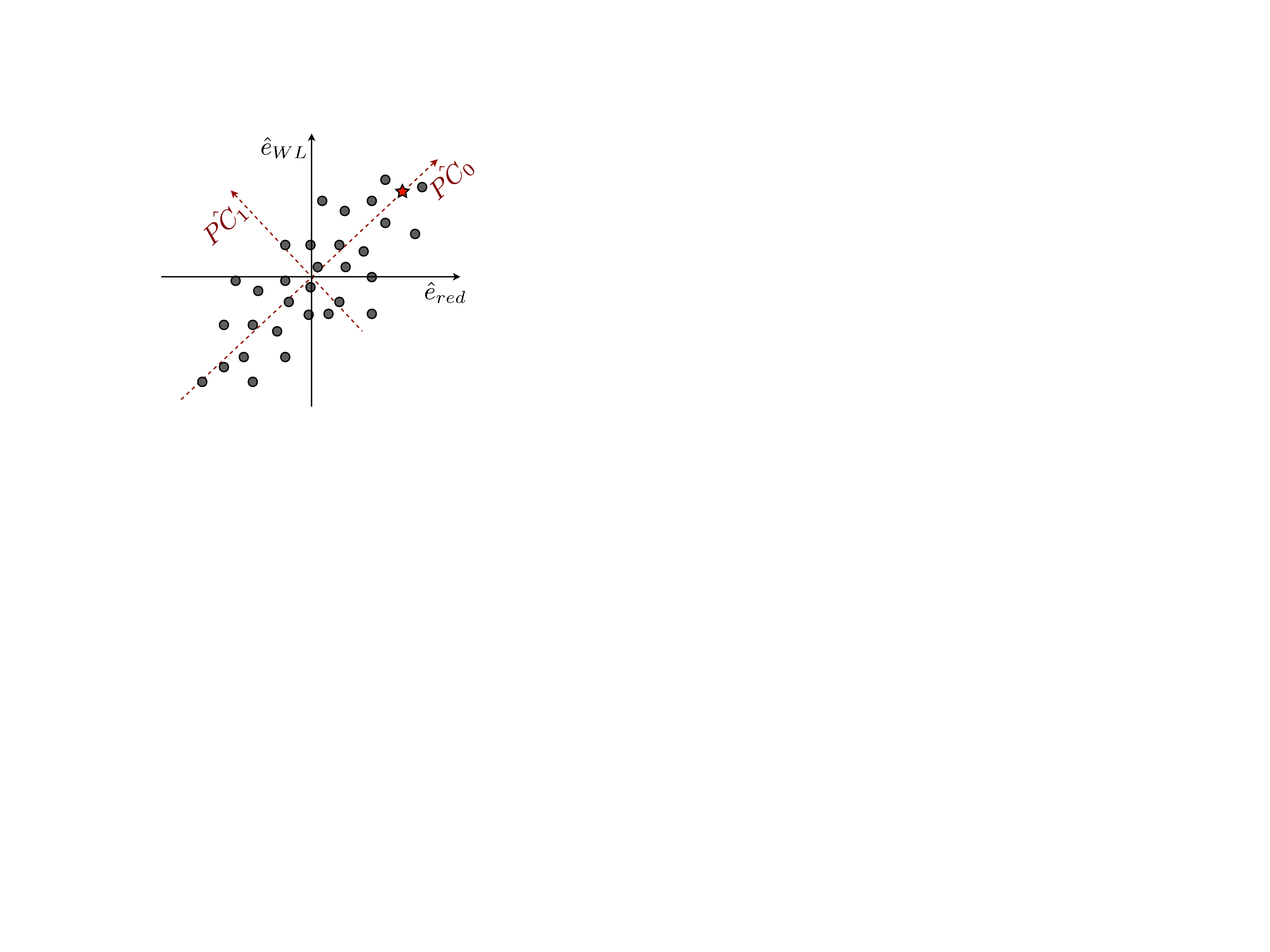}}
\end{center}
\caption{Principal component analysis takes a set of correlated
  measurements (using two methods shown as basis elements $\hat{e}_{\rm red}, \hat{e}_{WL}$) and
rotates their coordinates to 
a new basis, shown as dotted lines, where the measurements are
uncorrelated. 
For example, the point marked as a star (i.e.\ $\alpha = \star$) is 
$\vec{M}^\star = 10^{14} \hat{e}_{\rm red} + 10^{14} \hat{e}_{\rm WL}
= \sqrt{2}\times10^{14} \hat{PC}_0 +  0\; \hat{PC}_1$.
The generalization to more measurement methods and thus a higher
dimensional space is immediate.   Amongst the orthonormal
$\hat{PC}_i$, we  choose
$\hat{PC}_0$ to be along the direction of largest scatter 
(i.e. to correspond to the largest variance, $\lambda_0$), $\hat{PC}_1$ to be along the direction of second largest
variance,  etc., as described in the text.}
\label{fig:pcaexplain}
\end{figure}

For illustration,
we take a set hypothetical measurements for two methods, as shown in
Fig.~\ref{fig:pcaexplain}.  Each pair of measurements, by the two methods, is a position, i.e. a dot, in this plane labeled by two coordinates.
We take
one coordinate to be the shifted mass using red galaxy richness, $\mnred^\alpha=M^\alpha_{N_{red},
  est}- \langle  M_{N_{red},est}\rangle$, and the other to be the shifted
weak lensing mass, $\mwl^\alpha= M^\alpha_{\rm WL, est} -\langle M_{\rm WL, est}
\rangle$.   Here, $\alpha$ denotes which particular point is being
measured and the average is over all the points shown, all $\alpha$
values, i.e.  $\langle
M_{N_{red},est}\rangle =\frac{1}{N_\alpha}\sum_\alpha M^\alpha_{N_{red},est}$. The vector 
$\vec{M}^\alpha_{\rm obs}$ denotes $(\mnred^\alpha,\mwl^\alpha)$ and
$N_\alpha$ is the number of measurements (points) indexed by $\alpha$.  For our
first application below,  all different values of $\alpha$ pertain to
the same cluster, but label different lines of sight.
The shift 
by the average over all the points (all $\alpha$) guarantees 
that $\langle\mwl\rangle =
\langle \mnred \rangle = 0$.

Diagonalizing the covariance matrix for $(\mnred,\mwl)$, i.e. found by
summing over all $\alpha$, produces orthonormal 
eigenvectors $\hat{PC}_i$ (principal components) with eigenvalues
$\lambda_i$.  The eigenvectors are illustrated in
Fig.~\ref{fig:pcaexplain} and
are the axes of a new coordinate system in the space of measurement methods in which the measurements have zero
covariance.  
The $\hat{PC}_i$ 
can be expressed in terms of the original basis directions,
\begin{equation}
\hat{PC}_i = \beta_{\rm red, \; i} \hat{e}_{\rm red} +\beta_{\rm WL, \; i}
\hat{e}_{\rm WL} \; ,
\label{eq:betadef}
\end{equation} 
which identifies the mass scatters
contributing most to each $\hat{PC}_i$.
For example, if $\beta_{\rm red, \; 0}$ is large, then most of the scatter in the
direction of $\hat{PC}_0$ also lies in the direction of 
$\hat{e}_{\rm  red }$.  One also sees how different scatters are
related.  In this simple example
the biggest scatters come from increases or decreases in both
${M}_{\rm  red},{M}_{\rm  WL}$ simultaneously.  This implies, in this case,
that
$\hat{PC}_0$ might be related to an average overall mass shift,
amongst methods, as well.  Note that the 
overall signs for the $\hat{PC}_i$ are arbitrary.

A point (or vector from origin, labeled by $\alpha$) in the original space can
then be rewritten in the basis spanned by $\hat{PC}_0,\hat{PC}_1$:
\begin{equation}
\vec{M}^\alpha = M^\alpha_{\rm red} \hat{e}_{\rm red} +
M^\alpha_{\rm WL} \hat{e}_{\rm WL}
= a^\alpha_0 \hat{PC}_0 +  a^\alpha_1 \hat{PC}_1 \;.
\end{equation}
For example,
the point marked by a star in Fig.~\ref{fig:pcaexplain} 
has coordinates $\vec{M}^\star = 10^{14} \hat{e}_{\rm red} + 10^{14} \hat{e}_{\rm WL}
= \sqrt{2}\times10^{14} \hat{PC}_0 +  0\; \hat{PC}_1$.
That is,
$(M^\star_{\rm
  red},M^\star_{\rm WL})=(1,1)\times 10^{14}$ (assuming measurements are
in units of
$h^{-1} M_\odot$) and $(a^\star_0,a^\star_1)=(\sqrt{2},0)\times 10^{14}$.

The variances in each of the new directions, associated with the
coefficients $a^\alpha_i$, are the eigenvalues of the
principal components.  Thus, $\lambda_0$ is the eigenvalue associated with
$\hat{PC}_0$, etc.  For all PCA eigensystems we consider here, we will
order $i<j$ if  $\lambda_i > \lambda_j$ and define  $\sum{\lambda} =
\sum_i \lambda_{i}$ and $\prod \lambda = \prod_{i} \lambda_{i}$.
Generally, if there are $N_{\rm method}$ measurement methods, there are
$N_{\rm method}$
$\hat{PC}_i$, spanning an $N_{\rm method}$ dimensional space.  In
Fig.~\ref{fig:pcaexplain},
$N_{\rm method}$=2; when we consider the five different mass measurement methods below,
for example, $N_{\rm method}=5$.  For PCA applied to different cluster
properties, in \S \ref{sec:pca24} below, we
have $N_{\rm method}=24$.
We will apply PCA to covariance and correlation matrices, using
the Pearson covariance ${\rm Cov} (x y)  =\frac{1}{N_\alpha-1}
\sum_{\alpha=1}^{N_\alpha} (x_\alpha - \bar{x})(y_\alpha - \bar{y}) $ 
where $\bar{x}$ is the average of the $N_\alpha$ points $x_\alpha$, etc., and ${\rm Cov}(x,y)/\sqrt{{\rm Cov}(x,x) {\rm Cov}(y,y)}$ its associated
correlation, for $N_{\alpha}$ measurements.\footnote{
For the cluster mass scatters,
as this covariance can be affected by outliers,
we experimented with several outlier rejection schemes.  
 As the resulting
values were somewhat similar, and our analysis is in part just to
provide an example, we use the untrimmmed Pearson covariances and
correlations hereon.
We also
considered for some properties (as did some of the earlier cluster PCA
work) 
the Spearman correlation.
The Spearman correlation coefficient uses the ranking of the
measurements rather than the raw measurements themselves.  Trends were
similar to the Pearson covariances and correlations.
}

The sum of the $\lambda_i$, $\sum \lambda$, is the sum of the variances
of the measurement methods, their product, $\prod \lambda$, is related to the ``volume'' in this
space of scatters, i.e. how the measurements are spread out in the
space of
measurement methods  (specifically, for the example in
Fig.~\ref{fig:pcaexplain},  $\sqrt{\lambda_0 \lambda_1}$ is
proportional to the area of the ellipse).
A small eigenvalue $\lambda_i$ means that the scatter in the
corresponding $\hat{PC}_i$
direction is small, i.e. that the volume of scatters is roughly confined
to a lower dimension.   In particular,
if most of the scatter is due to
$\lambda_0$,
then there is a close
to linear relation present.  Such lower
dimensionality was used in \citet{Ein11} to come up with scaling
relations for superclusters. For PCA of correlations, a large $\lambda_0$ occurs if the initial
measurements via different methods
have strong correlations; for covariances, a large $\lambda_0$ can also
occur if an individual measurement method has large scatter.

Measurement methods $M^\alpha_{{\rm obs},j}$ (in the example $\mnred, \mwl$)
with the largest correlation or
covariance with $a^\alpha_i$ (their projection on
$\hat{PC}_i$) can be thought of as those dominating the scatter in the
direction of $\hat{PC}_i$. 
This covariance or correlation is not unrelated to the measurement method's contribution to $\hat{PC}_i$, 
($\beta_{{\rm obs}, j, i}$ in Eq.~\ref{eq:betadef}).  The covariance
or correlation 
is largest
when $\beta_{{\rm obs}, j, i}$ is large, and when the eigenvalue
$\lambda_i$ is large 
relative to the other $\beta_{{\rm obs},j}$'s and $\lambda_{i}$.  (For PCA on
correlations in particular, 
$\langle M_{{\rm obs},j} a_i\rangle/ \sqrt{\langle M_{{\rm obs},j}
  M_{{\rm obs},j}\rangle \langle a_i a_i\rangle} \sim
\beta_{{\rm obs},j, i}\lambda_i/\sqrt{\lambda_i \sum_k \lambda_k
  \beta_{{\rm obs},j,k}^2}$; if $\lambda_0, \beta_{{\rm obs}j,0}$ are
  large and $a_0$ is considered then this approaches 1.)

To summarize, PCA is a method taking a set of measurements via
different methods which are
correlated and separating them into uncorrelated combinations.   In
particular, if one denotes each set of measurements 
as a position in some space, along axes corresponding to
each type of
measurement method, then PCA is a rotation of coordinates in
this space.  The volume and shape traced out by the points representing the
measurements are related to the PC eigenvectors and eigenvalues, and
the
direction in the space corresponding to the largest PC eigenvector is
the combination of measurement methods with the largest scatter in its distribution.

\section{Variations for a single cluster due to line of sight effects}
\label{sec:indivar}

As mentioned in the introduction, many of the mass
measurement method scatters in
Fig.~\ref{fig:fullscatter} are 
comparable in size because they are due to similar properties of the
cluster or its environment.
Correlations between the scatters are thus expected, and as noted above,
these correlations and their consequences become increasingly
important as multiwavelength studies become more common.

We first consider each cluster and its line of
sight mass scatters separately.  In this way,
the ``true'' object and its true mass remained fixed; all variations in scatters are
due to changes in line of sight.
Several examples of correlations for these scatters were already noted and illustrated
in WCS\footnote{ An example of correlations between velocity dispersion and weak
lensing measurements are shown in WCS Fig. 14}, and correlations with various
physical properties (discussed below) were further 
studied in \citet{NohCoh11,Coh11}.  Here we statistically describe 
these correlations and
covariances.  We characterize
cluster to cluster trends and variations in line
of sight mass scatters and predicted masses, and then apply PCA to these scatters.

\subsection{Correlated mass scatters for different cluster observables}

\begin{figure}
\begin{center}
\resizebox{4.4in}{!}{\includegraphics{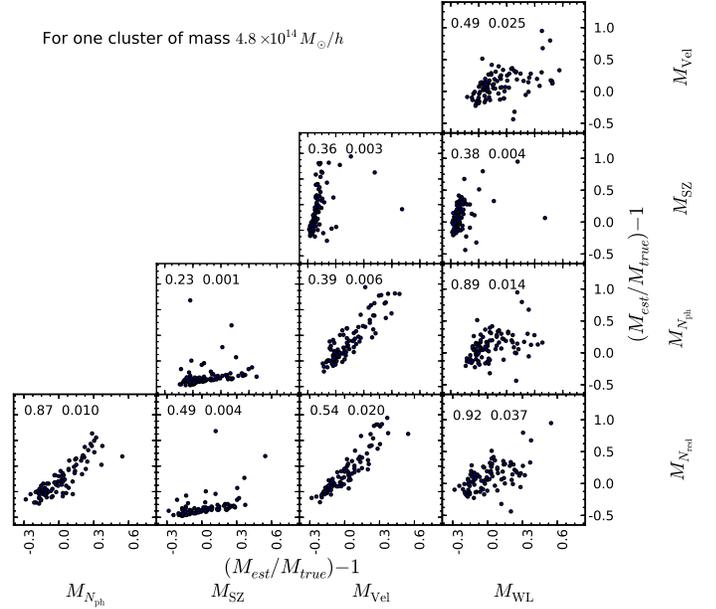}}
\end{center}
\caption{Cluster mass scatters ($M_{\rm
    est}/M_{\rm true} -1$) for
  one cluster of  mass
  $4.8 \times 10^{14} h^{-1} M_\odot$, along 95 lines of
  sight.  The correlation and covariance for each mass measurement
method pair is shown at the top of each panel (the y-axis is for a
larger
scale to allow room for these numbers).  Large correlations are
present for many pairs of mass measurement methods.
}
\label{fig:scatter9} 
\end{figure}

To give an idea of the correlations and covariances for our five
different mass measurement methods, we start with an example: the 10
pairs of
mass measurements for a single cluster $(M=4.8\times 10^{14} h^{-1}
M_\odot)$ shown in Fig. \ref{fig:scatter9}. 
Each panel shows a different pair of mass estimates along all lines of
sight (i.e. $N_\alpha = 95$), and correlations and covariances are listed at the top of each.
 We use $M_{\rm est}/\mtrue-1$ to
focus on fractional mass scatter.  As can be seen, many of the
correlations are large.

\begin{figure}
\begin{center}
\resizebox{4.4in}{!}{\includegraphics{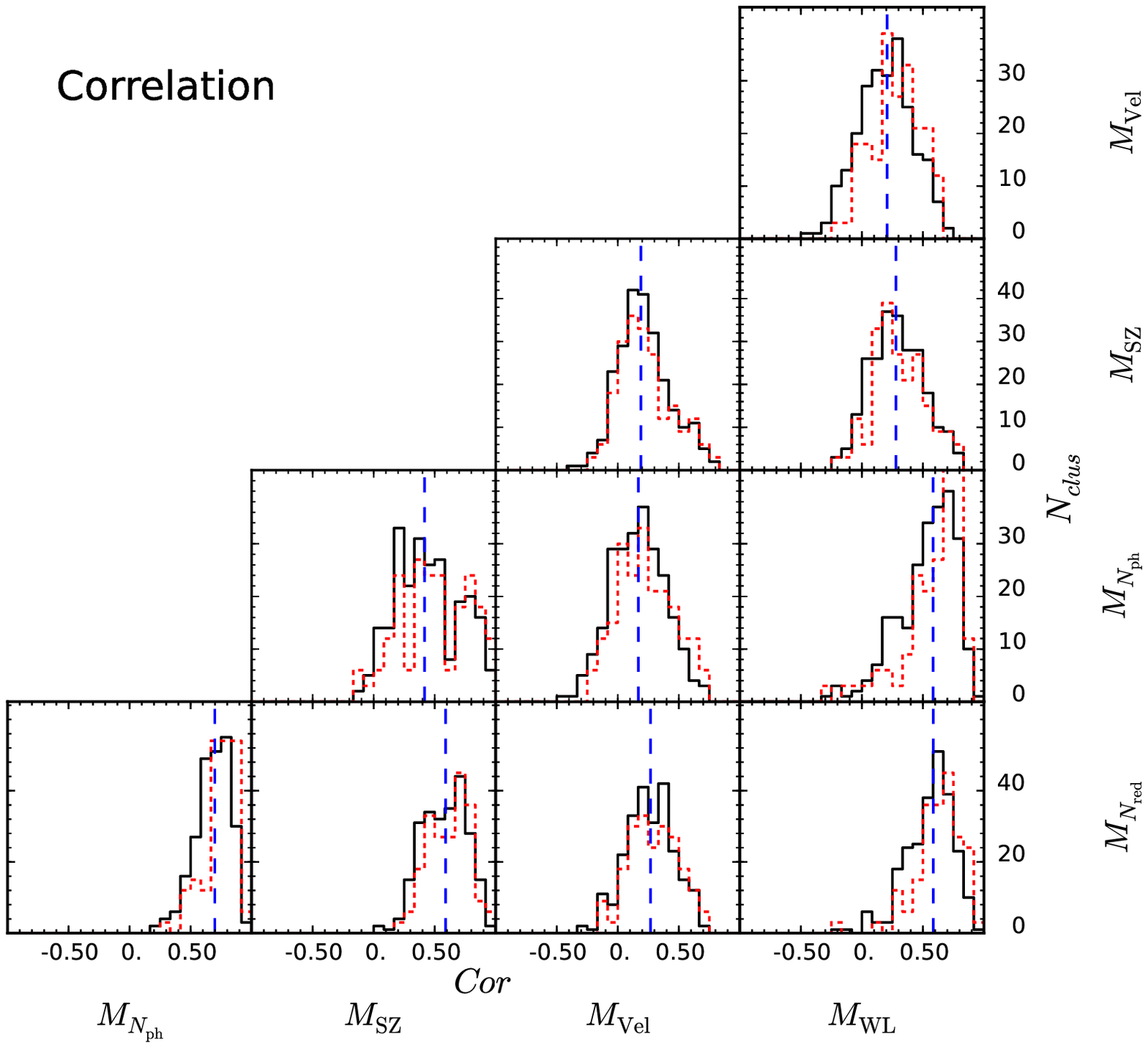}}
\end{center}
\caption{
Correlations of $M_{\rm est}/\mtrue-1$ for pairs of mass measurement methods
($\mnred$, $\mnph$, $\msz$, $\mvel$, $\mwl$). \newline
The solid line is the distribution of cluster mass scatter
correlations, for all clusters individually, for the same pairs as in Fig.~\ref{fig:scatter9}.
The dotted line corresponds to
the 70 clusters with mass $\geq 2\times 10^{14} h^{-1}M_\odot$. 
The vertical dashed lines are at the median values which are listed in
Table \ref{tab:corcov}, along with the average values.  Note that the x-axis,
the range of correlations, has a scale which varies widely between
different types of measurement method pairs.
}
\label{fig:corrscatter} 
\end{figure}

For all 243 clusters, 
the correlations for the same pairs of mass measurement methods are
compiled in Fig.~\ref{fig:corrscatter}, with medians and averages
given in Table \ref{tab:corcov}.
Strong correlations are frequent.
For each cluster, at least one pair of mass measurement methods has correlation $>$ 0.4,
and the largest pair correlation is often larger, $\sim 0.7$.
Within our cluster sample, the mass scatters for
($\mnred$, $\mnph$) are most often the highest correlated pair.  
The other measurement method pairs which frequently have the highest
correlation (but not as often
as ($\mnred$, $\mnph$)) are ($\mnred$, $\msz$), ($\mnred$, $\mwl$),
($\mnph$, $\msz$) and ($\mnph$, $\mwl$).
 The pair with the minimum correlation is most frequently
($\mvel$, $\mwl$) (closely
followed by ($\mvel$, $\msz$), 
and ($\mvel$, $\mnph$)).

\begin{figure}
\begin{center}
\resizebox{4.4in}{!}{\includegraphics{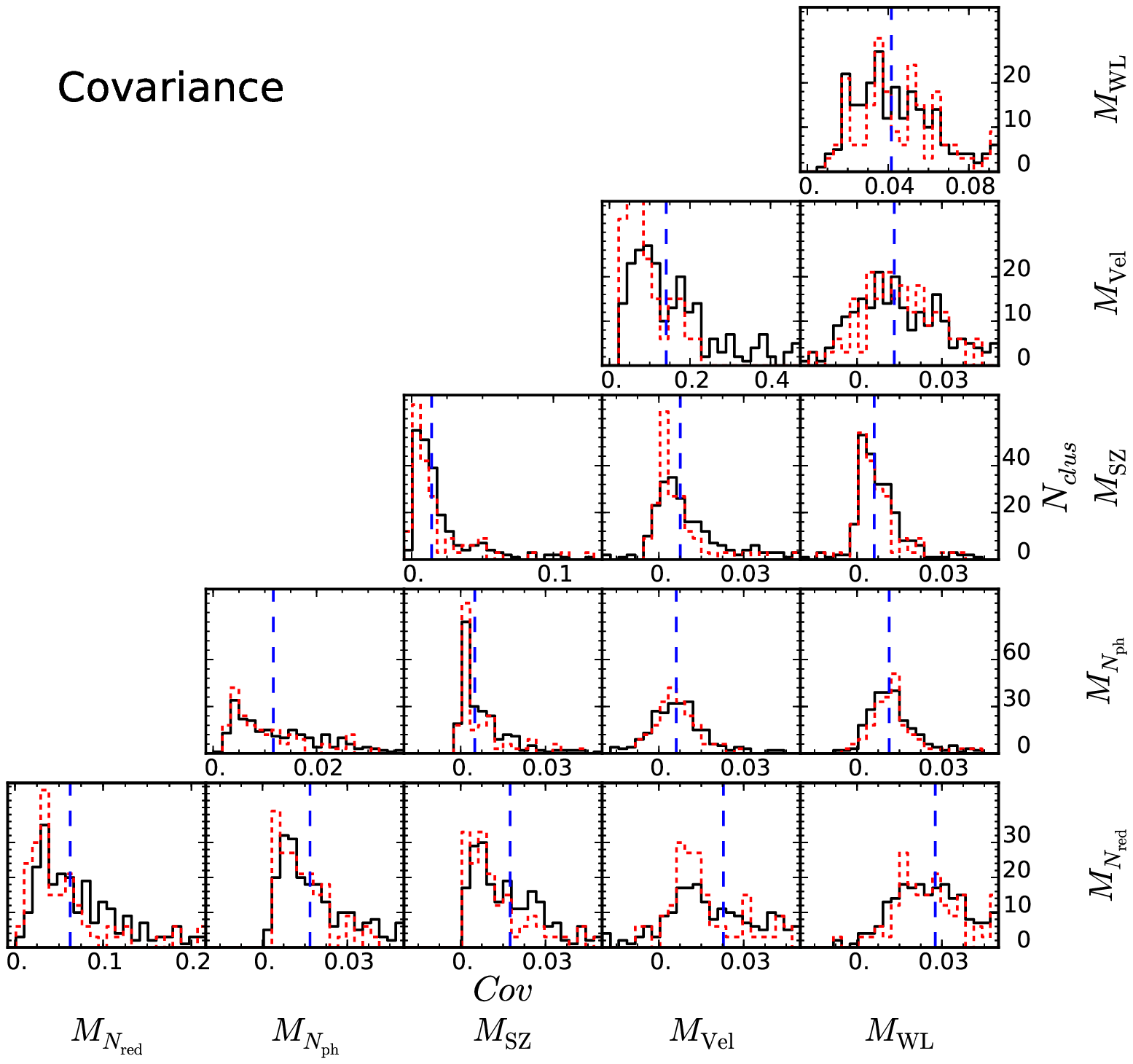}}
\end{center}
\caption{
Covariances of $M_{\rm est}/\mtrue-1$ for pairs of mass measurement methods
($\mnred,\mnph,\msz,\mvel,\mwl$). \newline
The solid line shows the distribution of cluster mass scatter
covariances, for all clusters individually, and the same mass
observation pairs as in Figs.~\ref{fig:scatter9},\ref{fig:corrscatter}.  The dotted line restricts to
the 70 clusters with mass $\geq 2\times 10^{14} h^{-1}M_\odot$.
As the characteristic mass scatter for most 
clusters is about 0.3, the range for covariances shown is
$\pm 0.05$.  The vertical dashed lines are at the median values, which
are also listed in Table \ref{tab:corcov}.
 Note that the x-axis,
the range of covariances, has a scale which varies with types of
mass measurement method pairs.
}
\label{fig:covscatter} 
\end{figure}

\begin{table}
   \centering
\begin{tabular}{lccccc} 
Pair  & Ave Cor & Med Cor& Ave Cov & Med Cov \\
$\mnred -\mnph$ &0.68   & 0.70  &0.025& 0.017\\
$\mnred -\msz$  &0.58   & 0.59 & 0.037 & 0.017 \\
$\mnred -\mvel$  & 0.26  &0.27  & 0.031 & 0.023 \\
$\mnred -\mwl$  & 0.55  & 0.58  & 0.031  & 0.028 \\
$\mnph -\msz $ &0.44   & 0.42   & 0.012 & 0.005\\
$\mnph -\mvel $ & 0.17  & 0.17   & 0.007 & 0.006 \\
$\mnph -\mwl $ & 0.53  & 0.58   & 0.013 & 0.011 \\
$\msz -\mvel$ &0.22  &0.19  &0.018 & 0.008 \\
$\msz -\mwl$ & 0.29  & 0.28  & 0.008& 0.006\\
$\mvel -\mwl$ & 0.19  & 0.21  &0.016  & 0.013 \\
$\mnred-\mnred$ & --  & --  & 0.084 & 0.062\\
$\mnph - \mnph$ &--  & --  & 0.015& 0.012\\
$\msz - \msz$ &--  & -- & 0.070 & 0.014 \\
$\mvel - \mvel$ & --  & --  & 0.205 & 0.142 \\
$\mwl - \mwl$ &--  & -- & 0.045 & 0.041 \\
\end{tabular}
   \caption{
Average and median values for the distribution
(for 243 clusters) of pairs of mass
measurement method correlations shown in
Fig. ~\ref{fig:corrscatter} and  covariances shown Fig. ~\ref{fig:covscatter}.
}
\label{tab:corcov}
\end{table} 

For our PCA analysis below, we will use covariances instead, shown in
Fig.~\ref{fig:covscatter}, with medians and averages given in Table \ref{tab:corcov}.
These are more relevant
for understanding the actual mass scatters and how they
change together, rather than, for example, how much a relatively large $\msz$ mass
scatter corresponds to a relatively large $\mnred$ mass scatter.  
As the fractional mass fluctuations (the
$\sigma$ values in Fig.~\ref{fig:fullscatter}) tend to be about 0.3,
the covariance sizes should be compared to $\sim$0.09.  The largest
covariances are between ($\mnred$,\ $\mwl$ and $\mvel$).
Each cluster has at least one covariance $\geq$
0.01. 
In contrast, the minimum covariance between measurement methods tends to be
between the pair ($\mnph$,\ $\msz$).\footnote{
If the abovementioned neglected weak lensing and SZ
scatter is uncorrelated with local properties of the cluster, as
expected, when this scatter is included the majority of the covariances will not change (except possibly the one
between
weak lensing and SZ, as they can possibly have correlated changes in
their scatter due to structure outside the box).
However, the correlations will change as they are divided by the
correlations of SZ with itself or weak lensing with itself.}  

Some of these correlation and covariance trends are understandable
(larger covariances
 tend to go with quantities with larger scatter more generally and
 vice versa), but others rely
upon the interplay between different measurement methods and the
causes of the scatter.  The relative importance of different
contributions to these were not a priori obvious to us, although
reasons could be found for trends.  For instance, the two richnesses
often
are the most correlated mass measurement pair.
This is perhaps because
both
use the same objects (galaxy counts), so that an enhancement
or decrement 
of $\mnph$ (galaxies making the spectroscopic cut) might be more likely accompanied by a
similar change in $\mnred$ (galaxies making the red sequence cut),
than by changes in the mass measurements by other methods.  
Other mass measurement methods are correlated with richness less
directly (for $\msz$ and $\mwl$ the measurements and thus presumably
the scatter are more directly
tied to the dark matter distribution rather than the biased galaxies;
$\mvel$ 
just seems to be weakly correlated with most things).  
In section \S \ref{sec:physdir} we compare the PCA results (scatters
which occur together) with
physical
cluster directions to get some idea of which properties might be
driving covariant mass measurement method scatters.

\subsection{PCA for individual clusters}
\label{sec:indiv_pc}
We now apply PCA to the covariances for $\vec{M}_{\rm obs}/\mtrue$ for
each individual cluster, to get a new basis,
\begin{equation}
\frac{\vec{M}_{\rm obs}}{\mtrue} =\frac{\vec{M}_{\rm
    est}^\alpha}{M_{\rm true}}-\frac{\vec{M}_{\rm ave}}{M_{\rm true}}  = \sum_i
a^\alpha_i \hat{PC}_{i,M} \; .
\label{eq:mobsdef}
\end{equation}
The subtracted offset $\vec{M}_{\rm ave} \equiv \langle \vec{M}_{\rm est} \rangle$,
where the average is over all the lines of sight for each cluster
of interest, i.e., there is a different $\vec{M}_{\rm ave}$ for each
cluster.  The median (and rms around zero) values for the ensemble of clusters, for 
$|\vec{M}_{\rm ave}/\mtrue - 1|$ are
[0.21 (0.32), 0.11 (0.17), 0.08 (0.15), 0.11 (0.18), 0.18 (0.27)] for
$\mnred, \mnph,\msz,\mvel,\mwl$ respectively.
The relative sizes of
the line of sight scatters around the average mass,
$\vec{M}_{\rm obs}/\mtrue - \vec{M}_{\rm ave}/\mtrue$ compared to
the line of sight averaged mass around the true mass
$\vec{M}_{\rm ave}/\mtrue - 1$ varies widely cluster to cluster.
Except for velocity dispersions,
the rms scatters of $(\vec{M}_{\rm
  obs}/\mtrue - \vec{M}_{\rm ave}/\mtrue)$ is $ \geq |\vec{M}_{\rm ave}/\mtrue
- 1|$ for 50-60 percent of the clusters
and the median value of $|\vec{M}_{\rm
  obs}/\mtrue - \vec{M}_{\rm ave}/\mtrue|$ is $\geq
|\vec{M}_{\rm ave}/\mtrue - 1|$ for about 30-40 percent of the clusters,
for velocity dispersions the numbers are closer to 90 and 70 percent
respectively.

As in section \S \ref{sec:pcadef}, and by the definition in Eq.~\ref{eq:mobsdef},
$M_{\rm  obs}$ refers to mass measurements which have zero
average when summed over the sample of interest.
Any vector in this space can of course be written in terms
of the orthonormal basis $\hat{PC}_{i,M}$; what is special for $\vec{M}_{\rm obs}$ is that
the variances of the $a^\alpha_i$ are equal to $\lambda_i$ for 
gaussian scatter.
Because there are five mass scatters,
there are five PC vectors $\hat{PC}_{i,M}$ per cluster, with
eigenvalues $\lambda_{i,M}$, again ordered $\lambda_{0,M}>\lambda_{1,M}>\lambda_{2,M}$
and so on.   
We use the subscript $M$ to distinguish these PC vectors
from others which will be considered in section \ref{sec:ensemble}, and take
$M_{{\rm obs},i}$, with $i=0,1,2,3,4$, to correspond to
$(\mnred,\mnph,\msz,\mvel,\mwl)$ and similarly for $\hat{e}_{{\rm obs},i}$. 
We also have, as in Eq.~\ref{eq:betadef},
\begin{equation}
 \hat{PC}_{i,M} =\sum_{{\rm obs},j = 0}^4  \beta_{{\rm obs},j,i} \hat{e}_{{\rm obs},j} \; .
\end{equation}

\begin{figure}
\begin{center}
\resizebox{3.5in}{!}{\includegraphics{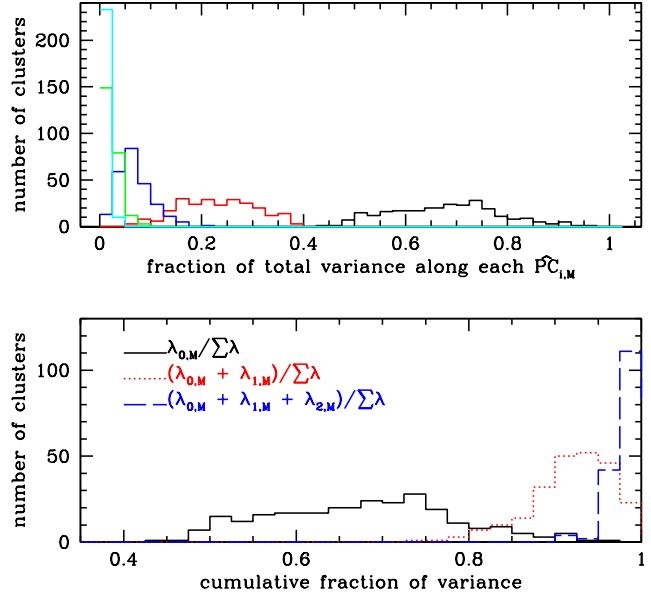}} 
\end{center}
\caption{(Top) Peaks from right to left:
fraction of covariance (i.e. scatter) $\frac{\lambda_{0,M}}{\sum{\lambda}}$,
$\frac{\lambda_{1,M}}{\sum{\lambda}}$, $\frac{\lambda_{2,M}}{\sum{\lambda}}$, etc.
On average $\frac{\lambda_{0,M}}{\sum{\lambda}}  \sim 0.7$.
(Bottom) Peaks from left to right: fraction of covariance from scatter
in
direction of $\hat{PC}_{0,M}, \hat{PC}_{0,M}$or $\hat{PC}_{1,M}$, etc.  Most of
the scatter is in the directions spanned by
$\hat{PC}_{0,M},\hat{PC}_{1,M},\hat{PC}_{2,M}$, with a substantial fraction
in the direction of the largest scatter.
}
\label{fig:pccontrib}
\end{figure}

We first consider the PC eigenvalues, the $\lambda_{i,M}$.
There are some trends:  the fractional scatter in the largest
direction ($\frac{\lambda_{0,M}}{\sum{\lambda}}$) $ \sim 0.7$, but can vary from 0.4 
to $\sim 1$, as shown in Fig.~\ref{fig:pccontrib}.
The relatively large contribution from $\lambda_{0,M}$
means that the variance is strongly dominated by the single combination of
mass scatters in the direction of $\hat{PC}_{0,M}$.
As seen in Fig.~\ref{fig:pccontrib}, bottom, $\lambda_{0,M},\lambda_{1,M},\lambda_{2,M}$ together
comprise almost all the variance for most clusters.  The presence of
some mass measurement methods with small scatter suggests that there
are 
some directions of the combined measurement methods which would also have
small scatter and thus small $\lambda_{i,M}$, and this is seen in the
much smaller values of $\lambda_{4,M}$ and sometimes $\lambda_{3,M}$.  The distribution of covariance
matrices shown in Fig.~\ref{fig:covscatter} determine the $\lambda_{i,M}$ 
when combined with the relation of the covariances to each other,
cluster by cluster.
The combination of measurement methods in $\hat{PC}_{4,M}$ which
has the smallest variance, $\lambda_{4,M}$ in our case, is also
interesting, we return to this in \S \ref{sec:ensemble}.

\begin{figure}
\begin{center}
\resizebox{3.5in}{!}{\includegraphics{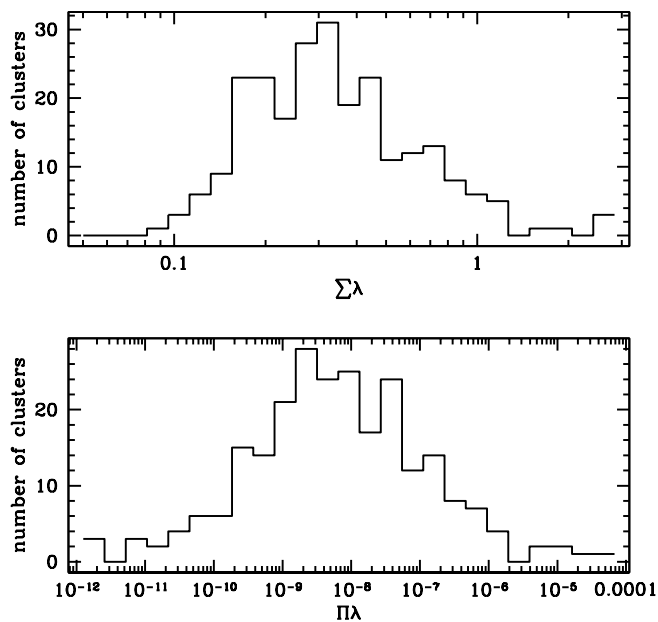}}
\end{center}
\caption{Sum and product of covariances for all clusters: the sum is
 dominated
by the eigenvalue in the direction of largest scatter, $\lambda_{0,M}$,
while the product, related to the volume in the space of scatters, can be made very small by small values of
e.g. $\lambda_{4,M}$.
The sum of eigenvalues peaks at $\sim 0.3$, the product peaks around
$10^{-8}$.  
}
\label{fig:variances}
\end{figure}

The sum and product of the $\lambda_{i,M}$ for all clusters
are shown on a logarithmic plot in
Fig.~\ref{fig:variances}.  
The sum of scatters can vary by a factor of
$\sim 30$ from cluster to cluster, and tends to be dominated by
$\lambda_{0,M}$, while
the product of the $\lambda_{i,M}$ can be made very small (its size varies by
over $10^7$) if some
directions, especially $\hat{PC}_{4,M}$, have very little scatter.
In \S \ref{sec:scalarprops} below, these scalar properties of the
cluster mass scatter will be compared
to physical cluster properties such as triaxiality and mass.

Turning to the PC vectors,
many clusters have similar $\hat{PC}_{0,M}$ (i.e. the combination of
mass scatters that dominates is similar for many of the clusters).
To quantify this more generally, we took $\hat{PC}_{i,M,minsq}$ as the direction
which minimizes $(\hat{PC}_{i,M} \cdot \hat{PC}_{i,M,minsq})^2$ for
the full ensemble of
243 clusters ($\hat{PC}_{0,M,minsq}$ is shown in the first line of Table \ref{tab:pcvecs}).
Fewer than 20 percent of the clusters have their $\hat{PC}_{0,M}$ pointing
more than $45^\circ$ away from $\hat{PC}_{0,M,minsq}$; about 25
percent have
their $\hat{PC}_{1,M}$ pointing more than $45^\circ$ away from
$\hat{PC}_{1,M,minsq}$. We 
found that for almost all the clusters the projection upon
$\hat{PC}_{1,M,minsq}$ is close in size to the projection on
$\hat{PC}_{0,M,minsq}$, 
(i.e. 0.62  correlation).  Continuing to $\hat{PC}_{4,M}$, the direction of least scatter,
$\sim$ 70 percent of the clusters are within $45^\circ$ of $\hat{PC}_{4,M,minsq} $
given in Table \ref{tab:pcvecs}.  This minimum scatter direction is
(not surprisingly) dominated by the $\hat{e}_{\rm N_{ph}},\hat{e}_{\rm
  SZ}$ directions, since these are the mass measurement methods with the
smallest scatter in our sample.\footnote{Again recall these particular
  coefficients 
do not
carry over directly to observations, most importantly because they are based on
scatter in $P(M_{\rm est}|\mtrue)$ and not vice versa.}
The similar forms of the $\hat{PC}_i$
suggest that the mass scatter combinations they correspond to might have similar
physical origins. 
For 75 percent of the
clusters, the coefficients of $\hat{PC}_{0,M}$ are also all the
same sign, that is that the dominant combination of scatter also has
all
the scatters increasing together relative to their average values, but 16/243 have large ($<-0.1$) opposite
sign coefficients for some mass measurement methods. 
As a large fraction of the variance is captured by $\hat{PC}_{0,M}$,
as seen in the large values of $\lambda_{0,M}/\sum \lambda$, the
coefficient of $\hat{PC}_{0,M}$ for the any given line of sight is
indicative of the size of scatter from the average along that line of sight.
\begin{table}
   \centering
\begin{tabular}{lccccc} 
Directions  &$\hat{e}_{\rm N_{\rm red}}$& $\hat{e}_{\rm N_{\rm ph}}$
&$\hat{e}_{\rm SZ}$ &$\hat{e}_{\rm vel}$ &$\hat{e}_{\rm WL}$ \\ 
\; \; \; $\hat{PC}_0$ & &  &  & &  \\
$\hat{PC}_{0,M,minsq}$&0.42 &0.14 &0.19 &0.83 & 0.29 \\ 
$\hat{PC}_{0,M,total}$&0.52 &0.21 &0.31 &0.69 &0.34 \\ 
$\hat{PC}_{0,M,total,massive}$&0.54 &0.25 &0.38 &0.51 &0.49 \\
\; \; \; $\hat{PC}_4$ & & &  &  & \\
 $\hat{PC}_{4,M,minsq}$&0.15 &-0.96 &0.2 &-0.01 &0.15 \\ 
$\hat{PC}_{4,M,total}$&0.32 &-0.94 &-0.02 &0.006 &0.098 \\ 
 $\hat{PC}_{4,M,total,massive}$&0.34 &-0.94 &0.03 &0.02 &0.06 \\
\end{tabular}
   \caption{
A comparison between
different characteristic directions in the space of mass scatters,
($\hat{e}_{\rm N_{\rm red}}$, $\hat{e}_{\rm N_{\rm ph}}$,
$\hat{e}_{\rm SZ}$, $ \hat{e}_{\rm vel}$, $\hat{e}_{\rm WL}$).
The direction which minimizes the
dot product squared with $\hat{PC}_{i,M}$ of all 243 clusters is
$\hat{PC}_{i,M,minsq}$.
$\hat{PC}_{i,M,total}$ are the vectors found by PCA analysis of all
observed cluster masses (normalized by $\mtrue$)
and all lines of sight, jointly, discussed in \S \ref{sec:ensemble};
$\hat{PC}_{i,M,total,massive}$ restricts to the 70 clusters
with
$M \geq 2 \times 10^{14} h^{-1} M_\odot$.
These vectors are derived from our estimates of $M_{\rm est}(\mtrue)$,
which
as mentioned earlier, neglects some systematics for WL and SZ and relies upon our
simulation calibrated mass
definitions in terms of observables.
}
\label{tab:pcvecs}
\end{table}

\begin{figure}
\begin{center}
\resizebox{3.5in}{!}{\includegraphics{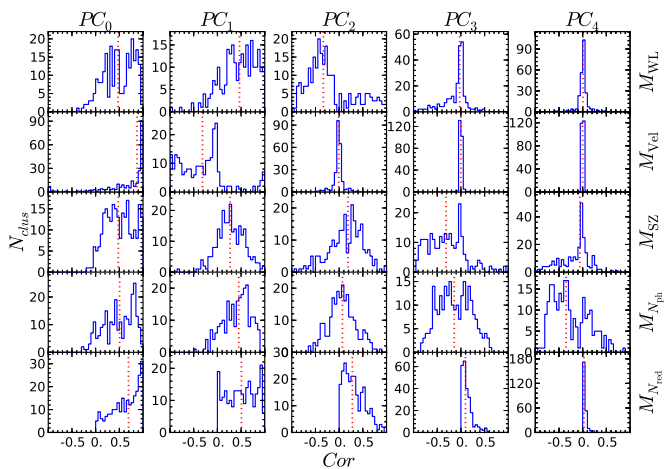}}
\end{center}
\caption{The correlations between the observed mass scatters and their projected
  values on $\hat{PC}_{i,M}$.
Velocity dispersions are very strongly correlated with the direction
of the largest scatter, $\hat{PC}_{0,M}$ in part because the component of
$\hat{PC}_{0,M}$ in the velocity dispersion direction, $\beta_{vel, \; 0}$,
tends to be large.  The dotted line marks the median value of the
correlations.
Note that the y-axis scale varies with mass measurement method and
$\hat{PC}_i$,
the x-axis does not.
}
\label{fig:corrmpc}
\end{figure}

The correlations of the five mass measurement methods $M_{{\rm obs},i}$
with the $\hat{PC}_j$ coefficients $a^\alpha_j$,
for all clusters, are shown in
Fig.~\ref{fig:corrmpc}.
For the ensemble of cluster measurements, 
the largest correlation with $a^\alpha_0$, i.e. with the coefficient
of $\hat{PC}_{0,M}$, is for velocity dispersions (in part because most
of the scatter is due to velocity dispersions).  
Correlations of $a^\alpha_0$ with
the other mass measurement methods are relatively smaller, and of similar
size to each other.  Taking instead the fraction of the mass scatter
vector due to $\hat{PC}_0$,  i.e.,  $a^\alpha_0/\sqrt{\sum_i
  (a^\alpha_i)^2}$, the correlations are weaker.  
In addition,  weaker correlations arise with the
coefficients of $\hat{PC}_{1,M}$, and for the direction with the
smallest direction of scatter, $\hat{PC}_{4,M}$, only $\mnph$ has a
noticeable correlation on average.

The direction of $\hat{PC}_{0,M}$ can also be compared to that of 
$\vec{M}_{\rm ave}/\mtrue$: the
average inner product
$\sim 0.7$, but the  peak is closer to 0.8, and there is a broad range
of values.
(The rest of the vector
$\frac{\vec{M}_{\rm ave}}{\mtrue}$ seems to lie in the
$\hat{PC}_{1,M},\hat{PC}_{2,M}$ 
plane).  That is, the direction of largest mass measured is
close to that of largest scatter; presumably this is because
this direction has large scatters generally.

We now compare the above quantities to line of sight dependent cluster
properties.  We will follow this in section \S \ref{sec:scalarprops} with
properties depending on the entire cluster rather than a given line of
sight.

\subsection{Relation to cluster line of sight properties}
\label{sec:physdir}
To get more understanding of the PCA decomposition, we compare values
of PC quantities along lines of sight to cluster
properties along those lines of sight.
The PCA line of sight dependent properties we consider are
the coefficients $a^\alpha_i$ of the $\hat{PC}_{i,M}$,
$a^{\alpha}_i/\sqrt{\sum_j (a^{\alpha}_j)^2}$ (the fraction of scatter in
different $\hat{PC}_i$ directions, $i$ and $j$ each run from 0 to 4),
and the total scatter for a given line of sight ($\sum_i
(a^{\alpha}_i)^2/\lambda_{i,M}$).  We correlate  these with the
angle $\theta_{\rm obs}$ between the line of sight and six specific physical
cluster directions, listed below.  

Previously, for this simulated data set, increased mass scatter was found by observing along certain
physical cluster directions by WCS; \citet{NohCoh11,Coh11}.  Our
extension using PCA includes six physical directions:
the long axis $\hat{\ell}$ of the cluster, calculated using the dark matter particles
in the simulation with a FoF finder as mentioned above, the plane normal
containing the most halo mass  $\hat{n}_{\rm mass}$, or connected filament
mass $\hat{n}_{\rm fil}$ centered on the cluster (see \S \ref{sec:filgroupdef} and 
\citet{NohCoh11} for more details), the direction of the largest
subgroup of galaxies $\hat{r}_{\rm sub}$ which originated from the
same infall host halo
(see \S \ref{sec:filgroupdef}),
and the velocity direction $\hat{v}_{\rm sub}$ of this largest subgroup.\footnote{ 
We also measured correlations with another direction dependent quantity, the amount of
substructure found via the Dressler-Schectman \citep{DreShe88} test, as
described
in \citet{WCS,Coh11}; however,
the correlations with $\hat{PC}_{0,M}$ were much weaker with this
directional dependent quantity than with the ones reported here.}
Many of these special cluster directions are similar to each other,
e.g. \citet{KasEvr05,WCS,Coh11},
as expected.  

\begin{figure}
\begin{center}
\resizebox{4.0in}{!}{\includegraphics{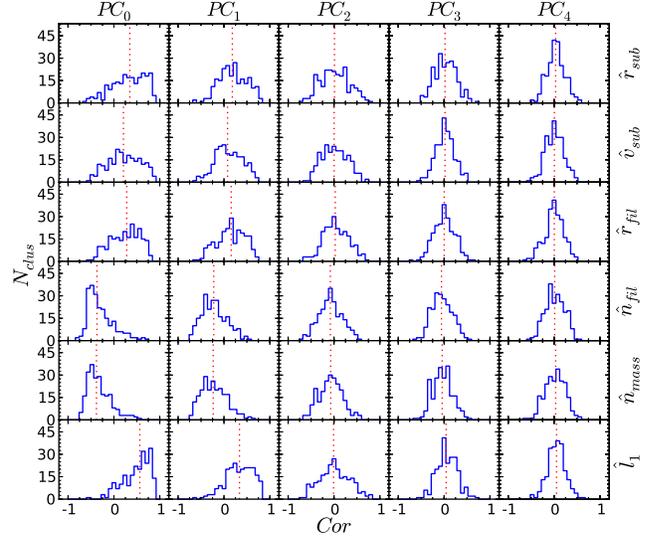}}
\end{center}
\caption{Correlations between 
the fraction of the line of sight 
mass scatter in the
$\hat{PC}_{0,M}$ direction and $|\cos \theta_{obs}|$.  The cluster
physical directions are $\hat{\ell}$, the long axis of the cluster,
$\hat{n}_{\rm mass}$, the direction perpendicular to the mass disk
with radius 10 $h^{-1} Mpc$ and width 3 $h^{-1} Mpc$ centered on the
cluster, containing the majority of the mass in halos, $\hat{n}_{\rm
  fil}$, similar to $\hat{n}_{\rm mass}$ in shape and volume, but
oriented to contain the majority of the
filamentary mass ending on the cluster, $\hat{r}_{\rm fil}$, the direction to the
most massive halo filament surrounding the cluster, and $\hat{v}_{\rm sub},\hat{r}_{\rm
  sub}$, the relative velocity and
direction of the largest galaxy subgroup in the cluster.
See text and \S \ref{sec:filgroupdef} for more
detailed definitions.
Only the 227 clusters which are filament endpoints are included in the
comparisons concerning filament planes. 
Vertical dotted lines denote medians, and the averages and medians are
shown in Table \ref{tab:vecprops}.
 The largest
contributions to $\hat{PC}_0$, for most clusters, seem to come when
observing along the long axis of the cluster, but there is a wide
scatter.   The correlations with the mass and filament planes are
maximized for observations along the planes, i.e. perpendicular to the
normal vectors $\hat{n}_{\rm mass}, \hat{n}_{\rm fil}$, as expected.
}
\label{fig:vecprop}
\end{figure}

The correlation of $a^{\alpha}_i/\sqrt{\sum_j
  (a^{\alpha}_j)^2}$ (the fraction of the 
mass scatter in the
$\hat{PC}_{i,M}$ direction, for the line of sight indexed by $\alpha$) with $|\cos \theta_{obs}|$ for each
observation is shown in Fig.~\ref{fig:vecprop}.
The medians and the averages of these correlation coefficient
distributions are shown in Table \ref{tab:vecprops}.
The largest average correlations of angles with the physical cluster axes are
with $\hat{PC}_{0,M}$, the direction of the combination of mass
scatters that dominates the scatter.  The average correlations with
$\hat{PC}_{1,M}$ are slightly smaller, for the rest of the
$\hat{PC}_{i,M}$ the average correlations tend to zero (as can be
seen, the individual clusters can have larger correlations).
For $\hat{PC}_{0,M}$, the largest correlation is with the direction of the
cluster long axis $\hat{\ell}$.  The next largest signals are with the direction
of the mass plane normal $\hat{n}_{\rm mass}$, 
filament plane normal $\hat{n}_{\rm fil}$, and the direction of the
largest substructure $\hat{r}_{\rm sub}$. 
The velocity of the largest substructure $\hat{v}_{\rm sub}$ and the direction of the most
massive filament $\hat{r}_{\rm fil}$ are more weakly correlated.  That is,
scatter dominated by the mass scatter combination in $\hat{PC}_{0,M}$
tends to occur more often when
the direction of observation is more aligned with the long axis of the
cluster.  (As the other cluster directions
are not linearly independent, strong correlations with them are possible and
seen as well.)
As most of the scatter occurs along $\hat{PC}_{0,M}$ and
$\hat{PC}_{1,M}$, both of which are most correlated with looking along the long axis,
it suggests that most of the scatter is due to looking
along the long axis $\hat{\ell}$.

\begin{table*}
   \centering
\begin{tabular}{cccccc} 
  & $\hat{PC}_{0, M}$ & $\hat{PC}_{1, M}$ & $\hat{PC}_{2, M}$ & $\hat{PC}_{3, M}$ & $\hat{PC}_{4, M}$ \\
$\hat{r}_{sub}$ & 0.35(0.31) & 0.18(0.19) & 0.01(0.02) & 0.02(0.03) & 0.03(0.04)\\
$\hat{v}_{sub}$ & 0.21(0.22) & 0.08(0.10) & 0.00(0.01) & 0.02(0.02) & 0.01(0.01)\\
$\hat{r}_{fil}$ & 0.28(0.24) & 0.16(0.15) & 0.03(0.04) & 0.00(0.01) & 0.00(0.00)\\
$\hat{n}_{fil}$ & -0.37(-0.29) & -0.22(-0.21) & -0.08(-0.08) & -0.06(-0.04) & 0.01(0.02)\\
$\hat{n}_{mass}$ & -0.38(-0.32) & -0.24(-0.21) & -0.07(-0.07) & -0.04(-0.04) & 0.03(0.02)\\
$\hat{l}_1$ & 0.56(0.50) & 0.34(0.33) & -0.01(0.02) & 0.04(0.06) & 0.05(0.05)\\
\end{tabular}
   \caption{The median of correlation coefficients between 
   the fraction of the line of sight mass scatter covariance in each $\hat{PC}_{i, M}$ direction and 
   $|\cos\theta_{obs}|$. The average is shown in the parenthesis. The full distribution is shown in 
     Fig.~\ref{fig:vecprop}. 
}
\label{tab:vecprops}
\end{table*} 
The correlation with $a^\alpha_0$, the full contribution from
$\hat{PC}_{0,M}$, rather than the fractional contribution above, was much weaker.

\begin{figure}
\begin{center}
\resizebox{3.in}{!}{\includegraphics{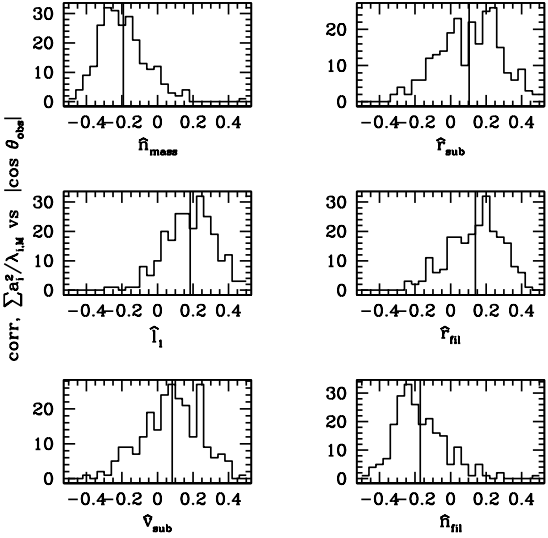}}
\end{center}
\caption{
Correlations of 
$\sum_i (a^{\alpha}_i)^2/\lambda_{i,M}$
(the weighted scatter of 
$M^\alpha_{\rm obs}$) with angle between
line of sight and various cluster axes, as described in the text and
in Fig.~\ref{fig:vecprop}.
Vertical lines are at average value.}
\label{fig:vecprop2}
\end{figure}

One other line of sight quantity, $\sum_i (a^{\alpha}_i)^2/\lambda_{i,M}$
(the
weighted scatter of 
$M^\alpha_{\rm obs}$), 
also tends to have correlations with special cluster directions
as seen in Fig.~\ref{fig:vecprop2}.  
The largest fraction of clusters with
correlations $> 0.20$ occurs with the long axis.\footnote{Correlations with the cluster long axis and external
  filaments or nearby clusters, both possible causes of mass scatters,
  have been
seen in many other works too, e.g.  \citet{vanBer96, Spl97, Col99,
  ChaMelMil00, OnuTho00, Fal02, van02, HopBahBod04, BaiSte05, Fal05,
  KasEvr05, Bas06, LeeEvr07, Lee08, PazStaPad08,PerBryGil08,RagPli07,
  CosSodDur10,CecPazPadLam11,PazSgrMerPad11}, correlations with 
the long axis and mass scatters specifically have been discussed
recently in,
e.g., \citet{BecKra11,Mar11,BatBonPfrSie11,BahMcCKin12,FerHob12}.}

To summarize this section,
each cluster's line of sight mass scatter variations and
correlations were analyzed
and characterized separately using PCA.  Both similarities
(similar large scatter directions) and differences (raw size of
scatter, amount of variation with line of sight) were found.  
The scatter was usually dominated (large $\lambda_{0,M}/\sum{\lambda}$) by one 
combination of mass scatters, which also tended to become
an increasing component of the total mass scatter as the angle of
observation became more aligned along the cluster's
long axis.

\section{Cluster to cluster variations}
\label{sec:scalarprops}
 Now we turn to
line-of-sight independent properties, PCA related or otherwise.
The scatters of each cluster are also characterized by several
numbers,
i.e. scalars,
which are not dependent upon the line of sight of observation
(e.g. $\lambda_0$).
Here we take some characteristic mass scatter scalars for each
cluster and compare them with other cluster properties,
intrinsic to the cluster or due to its environment, combining to give
24 quantities in total.
We make use of several quantities obtained previously (WCS;
\citet{NohCoh11,Coh11}) for this simulation 
and described in \S \ref{sec:filgroupdef}.  We use 
correlations rather than covariances, to take out the dimensional
dependence.  After considering properties of the pairwise
correlations in \S \ref{sec:corr24}, we consider the full ensemble of
correlations using PCA in \S \ref{sec:pca24}.

\subsection{Cluster quantities}
\label{sec:proplist}
For each cluster we consider the following (note that mass scatters
are for $\vec{M}_{\rm obs} = \vec{M}_{\rm est} - \langle \vec{M}_{\rm
  est} \rangle =\vec{M}_{\rm est} -\vec{M}_{\rm ave}$):
\begin{itemize}
\item  $\Delta M_i \equiv ({M}_{ave,i}-M_{\rm true})/M_{\rm true}$, observed average mass offsets
  for each of the five observables considered earlier: red galaxy
  richness,
 phase richness, SZ, velocity dispersions and weak lensing
 respectively.  Note that these correlations are identical to those of
$M_{ave,i}/M_{\rm true}$.
\item  $|\cos \theta_{0,minsq}|$, i.e. $|\hat{PC}_{0,M} \cdot
\hat{PC}_{0,M,minsq}|$, the angle between the largest mass scatter direction for each
cluster, $\hat{PC}_{0,M}$, and the direction $\hat{PC}_{0,M,minsq}$, given in Table \ref{tab:pcvecs},  which minimizes the dot
product with $\hat{PC}_{0,M}$ for all the clusters.
\item  $\sum{\lambda}$, the sum of the cluster's mass scatters in \S \ref{sec:indiv_pc}.
\item  $\prod \lambda$, the  product of the cluster's scatters in \S \ref{sec:indiv_pc}. 
\item  $\frac{ \lambda_{0,M}}{\sum{\lambda}}$ fraction of total (mass scatter) 
  variance along direction of largest variance in \S \ref{sec:indiv_pc}. 
\item  $\lambda_{1,M}$,  variance along the direction with second
  largest mass scatter variance in \S \ref{sec:indiv_pc}. 
\item  $\lambda_{4,M}$, variance along the direction with smallest mass
  scatter 
  variance in \S \ref{sec:indiv_pc}. 
\item  $T = \frac{l_{1}^{2} - l_{2}^{2}}{l_{1}^{2} - l_{3}^{2}}$,
  triaxiality, where the $l_1,l_2,l_3$ are the axes of cluster, calculated
  using the dark matter particles in the FoF halo.
\item  $S = \frac{l_{3}}{l_{1}} \ \ \ (l_{1} > l_{2} > l_{3})$,
  sphericity.
\item  $f_{M_{\rm fplane}}$, the fractional connected filamentary mass in the local plane
around clusters defined in \S \ref{sec:filgroupdef}.
\item  $f_{M_{\rm hplane}}$, the fractional halo mass in the local plane
around clusters, see \S \ref{sec:filgroupdef}.
\item  $M_{\rm sphere}$, the mass in halos above $5 \times 10^{13}
  h^{-1}M_\odot$ within a 10 $h^{-1} Mpc$ radius sphere of the cluster.
We used the sum of large halo masses in a 10 $h^{-1} Mpc$ radius sphere
around the 
central cluster rather than the total
mass because the former is already known to be correlated
with the mass of the central halo
(e.g., \citet{NohCoh11}).
\item  $\mtrue$, the cluster (FoF $b=0.168$) mass (also called $M$).
\item  $f_{R_{\rm sub}}$, 
the fractional richness of the largest galaxy subgroup (see \S \ref{sec:filgroupdef}).
\item  $f_{D_{\rm sub}}$,
the ratio of the distance to the largest galaxy subgroup in the
cluster to the length of
the longest cluster axis (see \S \ref{sec:filgroupdef}).
\item  $c_{\rm vir}$, the concentration (not scatter from mean
  concentration of a given mass), from
  fitting all of the Friends of Friends halo particles to an NFW
  \citep{NFW} profile.  This was found to be very important in the
  previous
studies \citep{Jee11,SkiMac11,Ein11} inspiring this work.
\item  $t_{1:3}$, the time of most recent $\geq$1:3 merger (often
  taken as the threshold for a merger-driven starbust)
\item   $t_{1:10}$, the time of most recent $\geq$1:10 merger (often
  taken as a threshold for merger-driven AGN feeding)
\item $x_{\rm off}$, the distance between the central galaxy position
  and the average of the galaxy positions.  This
  is
similar to the ``relaxedness'' considered by \citet{SkiMac11}, but in that
case they use the offset between
the most bound particle and the halo center of mass.  We
also considered scaling by $M^{1/3}$, the results did not change
significantly.
\item  $\cos \theta_{\Delta,0}$, the cosine of the angle between
$\hat{PC}_{0}$ and $\vec{M}_{\rm ave}$, i.e., the largest mass scatter
direction vs. the average mass offset direction.   
\end{itemize}

\begin{figure}
\begin{center}
\resizebox{4in}{!}{\includegraphics{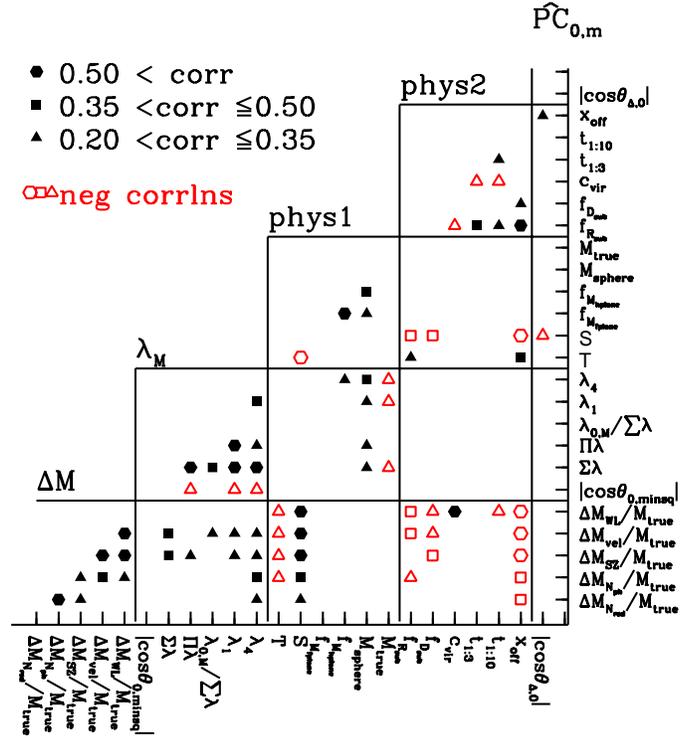}}
\end{center}
\caption{Correlations between cluster properties:
$0.5<corr$ (filled hexagons), $0.35<corr\leq 0.5$ (filled squares), $0.2
< corr \leq 0.35$(filled triangles), opposite signs are open versions
of the same symbols,
i.e. $corr<-0.5$ (open hexagon), etc.  Auto-correlations are not
shown. 
Cluster properties are
described in
section \ref{sec:proplist}.  
A blank space means that a correlation has
absolute value $\leq$ 0.2.   The horizontal and
vertical lines distinguish the types of cluster properties.  $\Delta
M$ labels the five components of $\Delta \vec{M}$.  $\lambda_M$ refers to
measurements for that cluster's mass scatter correlations.  ``phys1''
refers to cluster properties which are environmental or shape
related.  ``phys2'' refers to cluster properties more related to
substructure or
merging (concentration, distance to largest substructure, etc.),
$\lambda_0$
refers to dot product or correlation with $\hat{PC}_{0,M}$.
Only half of the correlations are shown, as they are symmetric
across the
diagonal axis.}
\label{fig:scalarpair}
\end{figure}

We thus have a space of dimension $N_{\rm method}=24$  
for the correlation and PCA analysis below.
Several of these quantities might be expected to be related.
For instance, sphericity and
triaxiality both characterize departures from perfect spheres, 
but triaxiality measures prolateness and oblateness 
while sphericity is sensitive to flatness.
Similarly the sum and product
of the eigenvalues probe the size of the largest eigenvalue and
in principle how the largest and smallest eigenvalue change together,
respectively.
A priori, it isn't clear which of our large set of
quantities have the strongest or most illuminating relations to
each other, so we start with a large set.

\subsection{Correlations}
\label{sec:corr24}
Fig.~\ref{fig:scalarpair} summarizes the pairwise
correlations.  For simplicity, only correlations
which have absolute 
value $>0.2$, in six ranges, $\pm 0.2, \pm 0.35, \pm 0.5$, are shown.  The correlations
of measurements with themselves are omitted.
Filled dark (open red) symbols are positive (negative) correlations.
Properties in the list are grouped by type: offsets
of  cluster
average mass measurements from their true mass 
(``$\Delta M$'', defined in \S \ref{sec:proplist}), PCA
related scalars for each cluster from \S \ref{sec:indiv_pc}
(``$\lambda_M$''), 
cluster environment or shape (``phys1''),
cluster history (``phys2''), and  $|\cos \theta_{\Delta,0}|$.
Relations between
measurements of the same sort (e.g. concentration and time of last
merger, etc.) can be seen in the diagonal boxes.
 
The off-diagonal boxes thus correlate different sorts of cluster
properties.  We concentrate on these.  The first thing to notice is
that 
many large correlations are visible. 
We start with
correlations
with $\Delta \vec{M}$, each cluster's five average fractional mass measurement
offsets, corresponding to the five measurement methods.  The different components of $\Delta \vec{M}$ often have similar correlations with other
quantities.  In particular, they are all strongly correlated with
sphericity, most are anticorrelated with triaxiality and all are
anticorrelated with the offset of the average galaxy position.  All
but the weak lensing mass offset are also correlated with the smallest
mass variance $\lambda_{4,M}$, that is, when the average mass offsets
$\Delta M_i$ are large, the variance in the direction of minimum
scatter around the average measured value tends to be large as well:
the line of sight averaged scatter often is increasing for clusters
with larger mass scatters around these averages.
There is also a trend of these average mass offsets being
anticorrelated with signatures of relaxedness (fractional richness of
the
largest subgroup, offset of the average galaxy position relative to
the
galaxy center, etc.).

Turning to other correlations with the mass scatter variances, an
increase in
the mass in halos around the cluster tends to be accompanied by an increase many of the
measures of mass scatter, i.e. properties associated with the
$\lambda_{i,M}$.  This might indicate the presence of a supercluster.
The size of the smallest mass scatter variance
$\lambda_{4,M}$ also seems to increase with increasing fractional
halo mass in the plane of the cluster, $f_{M_{hplane}}$.

Lastly, triaxiality is correlated with indicators of substructure,
(the offset of the galaxy average
position from the central galaxy and the fractional richness of the
largest galaxy subgroup), while in contrast sphericity is anticorrelated with
these, as well with the fractional distance of the largest galaxy
subgroup from the cluster center.  Physically, one might expect a rich
subgroup in the cluster to lie along the long axis of the cluster
(e.g. coming from a filament feeding matter into the cluster) and be
correlated
with a lengthened (more triaxial) cluster as a result.
\subsection{PCA}
\label{sec:pca24}
By using PCA on these pairwise correlations, more general groupings can be investigated.
This application of PCA to our sample is an extension of PCA 
of several cluster properties studied by
\citet{Jee11,SkiMac11} and by \citet{Ein11} for superclusters, which inspired
this study.  Properties correlated by
\citet{Jee11, SkiMac11} included virial mass, concentration, age,
relaxedness, sphericity, triaxiality, substructure, spin, and
environment.  \citet{Jee11} used 1867 halos from
several boxes with masses ranging from $ \sim 10^{11} h^{-1} M_\odot$
up and found that the dominant PC vector was most correlated with
halo concentration.  \citet{SkiMac11} used
several boxes with halos with masses $ \geq 10^{10} h^{-1}
M_\odot$, and also found halo concentration to be
important for clusters, as well as halo mass and degree of 
relaxedness.\footnote{For superclusters in SDSS DR7, \citet{Ein11}
  considered weighted luminosity, volume, diameter, density of the
  highest density peak of galaxies and the number of galaxies, as well
  as shape parameters, and used the two largest $\hat{PC}_i$ to find
  scaling relations.  They divided up the sample into two sets of
  superclusters based upon where they lay in the planes spanned by
  pairs of $\hat{PC}_{0},\hat{PC}_{1}, \hat{PC}_2$.}  Our sample is
closest to the high mass tail of these samples.

\begin{figure}
\begin{center}
\resizebox{3.7in}{!}{\includegraphics{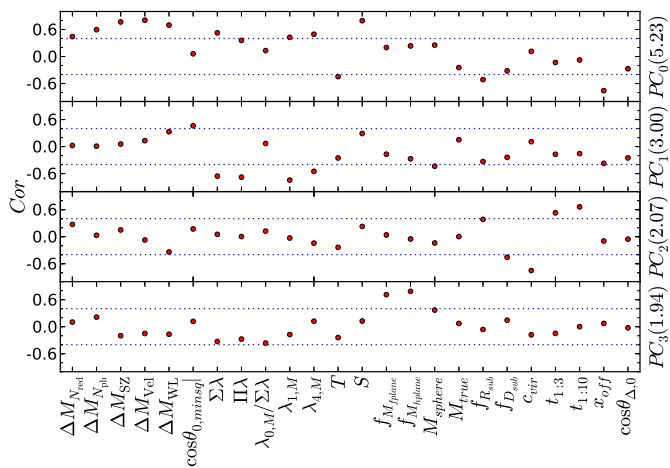}}
\end{center}
\caption{Correlation coefficients between the cluster properties
  listed in \S\ref{sec:proplist} and their projection on the first four principal components.
Horizontal dotted lines at $\pm 0.4$ are to guide the eye to larger
positive or negative correlations.
The top box is for $\hat{PC}_0$, the second is for $\hat{PC}_1$ and so
on; the expansion of the $\hat{PC}_i$ vectors is given by these
correlations divided by the $\lambda_i$ eigenvalues, 5.23, 3.0, 2.07, 1.94
respectively for $\hat{PC}_0,\hat{PC}_1,\hat{PC}_2,\hat{PC}_3$.}
\label{fig:allscalcor}
\end{figure}

The 24 PC eigenvalues\footnote{Some of the previous analyses consider logarithms of
the scalar quantities, we redid the analysis taking logarithms of
 $\sum{\lambda}$, $\prod \lambda$, $\lambda_{1,M}$, $\lambda_{4,M}$, $f_{M_{\rm
     fplane}}$, $f_{M_{hplane}}$, $f_{R{\rm sub}}$, $f_{{D_{\rm
       sub}}}$, $c$, $M_{\rm
   sphere}$, $ M$ 
and found very similar results for correlations with the first 2
$\hat{PC}_i$ as below, for $\hat{PC}_2$, correlations roughly increased for the merger related quantities, and decreased for the $M_{\rm sphere}$ and
plane environment quantities.}
are $\lambda_i/\sum{\lambda}$=(0.22, 0.12, 0.09, 0.08, 0.07, 0.05
(3 times), 0.04, 0.03 (4 times), 0.02 (3 times),  0.01 (6 times),
$<0.005$ (2 times)).
Unlike the case of mass scatter, where $\lambda_{0,M}$ is relatively large, 
(on average $\lambda_{1,M}/\lambda_{0,M}<0.4$), 
here $\lambda_1/\lambda_0 \sim 0.6$,
and $(\lambda_1,\lambda_2,\lambda_3,\lambda_4) = (5.23,3.00,2.07,1.95)$, fairly close to each other.
This makes interpretation less straightforward. 
However, subsets of properties with strong correlations with individual $\hat{PC}_i$
may indicate these properties change together.  To better identify relations,
one can take these subsets and do PCA on the correlations within this
subset alone.  An example is given below.
The expansion coefficients of the  $\hat{PC}_i$ for the different
measured quantities, the
$\beta_i$ in Eq.~\ref{eq:betadef},
are given by the correlations divided by the eigenvalues.
(For example, the expansion coefficients of the $\hat{PC}_{0}$ 
in the order listed in \S \ref{sec:proplist} are
(0.19, 0.26, 0.34, 0.35, 0.30, 0.03, 0.23, 0.16, 0.06, 0.19, 0.22, -0.20, 
0.35, 0.09, 0.10, 0.11, -0.11, -0.22, -0.14, 0.05, -0.06, -0.03, -0.33, -0.12).)

The strongest correlation is between the velocity dispersion mass
offset $\Delta \mvel$ and $\hat{PC}_0$ ($\sim$0.8), 9 other properties have
correlations with absolute value above or equal to 0.7 with at least one of the
$\hat{PC}_i$ ($i \leq 6$).  Considering smaller correlations, 21/24 of the properties
have at least one correlation of absolute value $\geq 0.5$ with one
of the $\hat{PC}_i$, $i \leq 11$, and all properties have at least one
correlation $\geq 0.4$ with at least one $\hat{PC}_i$. 
The correlations of the first four principal component vectors with
the 24 properties are shown in Fig.~\ref{fig:allscalcor}.
Lines are drawn at $\pm 0.4$ to guide the eye to the larger
correlations.
The sign of the correlations with the eigenvectors depends upon
the overall sign of the eigenvectors, which is arbitrary, but the
relative signs of the correlations of the different properties with
each eigenvector are not arbitrary.

We begin with $\hat{PC}_0$, the direction of largest variation.
The $\Delta M_i$'s, the
sum of the mass scatter variances $\sum{\lambda}$, the individual
mass scatter variances $\lambda_{1, M},\lambda_{4,M}$ and sphericity
$S$ all have
large correlations ($\ge 0.4$) with their projections onto
$\hat{PC}_0$.  These correlations are in the opposite sense of those of $T$, $f_{R_{sub}}$
and $x_{off}$ (these latter three might indicate less relaxed halos).
A relation between large
triaxiality and an increased offset in the average galaxy position or
the presence a rich
subgroup is somewhat intuitive, but the sign of the relation to the
average mass measured for the cluster is surprising.  One
interpretation of the relation between $\Delta M_i$ and $T$
seen both here and in the correlations is that clusters which are more
triaxial have fewer lines of sight along the elongated and presumably
large measured mass direction, so that fewer lines of sight result in
a large $\vec{M}_{\rm est}/\mtrue -1$.  See also related discussion in
\citet{Ras12}.\footnote{We thank E. Rasia for discussions about this.}

To go further, we considered the subset of properties 
which correlate
strongly with $\hat{PC}_0$ ($\Delta M_i$,$S$,$T$,
$f_{R_{sub}}$,$x_{off}$,$\sum\lambda$, $\lambda_{1,M}$,$\lambda_{4,M}$) and 
applied PCA to their intercorrelations.
The trends with $\hat{PC}_0$ for these quantities appear
with the direction $\hat{PC}_{0,{\rm subset}}$ as well.  In this
case, however, $\lambda_{0,{\rm subset}}$ is relatively larger, $\lambda_{0,{\rm subset}}/\sum \lambda= 0.39$.   To help in understanding, we took three groups of properties
(1) $\Delta M_i$ (except $\Delta \mwl$);
(2) the mass scatter measures $\sum \lambda,
\lambda_{1,M},\lambda_{4,M}$;
and (3) $T$,$f_{R_{Sub}}$,$x_{off}$, $S$, $\Delta \mwl$.  The
properties in (1) and (2) have large same sign correlations with
$\hat{PC}_0$.  For (3), sphericity $S$ and average
weak lensing mass offset $\Delta \mwl$ have this sign correlation as
well, while $T$, $f_{R_{Sub}}$,$x_{off}$ have the opposite sign.
These three groups (roughly) can be seen in correlations with
$\hat{PC}_{1,{\rm subset}}$ (0.17 of the sum of eigenvalues), but in this case the
relative behavior between properties in (2) and (3) reverses
(i.e. increased mass scatter now has the opposite correlation with
$\hat{PC}_1$ as increased sphericity does), and changes in (1) all
have correlations
$<0.07$ with $\hat{PC}_{1,{\rm subset}}$. (Note that the
general orthogonality of the $\hat{PC}_i$ means relations between
properties, i.e. relative sizes and signs of correlations,
do have to change from from one $\hat{PC}_i$ vector to another.)
The correlation strengths change
as well, going from $\hat{PC}_0$ to $\hat{PC}_1$, increasing for (2)
and
decreasing (below 0.4) for (3).
Considering the trends in broad brush, for a given cluster,
large measured average (over lines of sight) mass offsets from true
mass, aside from weak lensing, 
seem to tend to come with high sphericity and large mass measurement
scatters, and low triaxiality, $f_{R_{sub}},x_{off}$.  That is, when a
cluster's mass
offsets are very large, there is a tendency for the vector of all its
properties to be lying far along
$\hat{PC}_{0,{\rm subset}}$, implying trends for the other properties.
When average
mass offsets (besides $\Delta \mwl$) are small, it is not clear
whether a given cluster's variations are lying along $\hat{PC}_{0,{\rm subset}}$ or
$\hat{PC}_{1,{\rm subset}}$ or some other direction.  As the relation between
large sphericity (and small triaxiality, $f_{R_{sub}},x_{off}$), i.e., (3), 
and the size of mass measurement scatters, i.e., (2), reverses between these two
PCA vectors, small mass offsets makes it difficult to estimate
the relation between them (i.e., between (2) and (3)).  

Continuing with the full set of 24 properties, and considering correlations with $\hat{PC}_1$,
the large sphericity/small triaxiality, (relatively) small richness in
biggest subgroup, (relatively) small distance of largest subgroup from
center and small offset of average galaxy positions with
central galaxy position relation trends are also seen, but more weakly.
Fluctuations in the direction of
$\hat{PC}_1$ have an increase in
$|\hat{PC}_{0,M} \cdot \hat{PC}_{0,M,minsq}|$ tied to a decrease in overall mass scatter ($\sum
\lambda,\prod \lambda, \lambda_{1,M},\lambda_{4,M}$) and a decrease in mass in
nearby large halos $M_{\rm sphere}$.  This last perhaps indicates that
the
alignment of $\hat{PC}_{0,M}$
is better, and the mass scatter smaller, when the cluster is not in a 
supercluster.  
Low concentration and relative distance of the largest subgroup from the cluster center are
accompanied by earlier major mergers for variations in the direction of $\hat{PC}_2$.
 The fraction of mass in
the filament and mass planes seem to change together for changes along
$\hat{PC}_3$, along with the amount of mass in nearby large halos.
Again, these last 3 properties can be studied alone via PCA.  In this case the associated
$\lambda_{0,{\rm subset}}/\sum \lambda=0.65$.  Note this is a different
subset of PC vectors than considered earlier, and corresponds to all
three quantities,
normalized by covariance, changing by about the same
amount with the same sign.  
This may imply that clusters which are in a richer environment, that is, with more
nearby halos, may have the mass around them distributed in a
more
planar shape.

We experimented with a larger set of
properties than shown here, 
including for example
the quantities shown in the first column of Fig. ~\ref{fig:vecprop} for each cluster,
i.e. the correlations with projections on $\hat{PC}_{0,M}$ and
physical directions.   In this case
the correlations with $\hat{PC}_0$ for the other quantities remained essentially as shown in the top line of 
Fig.~\ref{fig:allscalcor} but 
 the $\hat{PC}_{0,M}$-long axis projections and the $\hat{PC}_{0,M}$-direction of largest subgroup
projections had large correlations with $\hat{PC}_0$ in the same sense
as $T$, cluster triaxiality.  

\section{An ensemble of clusters}
\label{sec:ensemble}
The previous sections considered cluster directional mass scatters and
variations in properties, cluster to cluster.  One can also consider
the joint ensemble of mass measurements of all the simulated clusters
and all their lines of sight, for all five methods. 
A comparison is interesting between the PC vectors of the
individual clusters, for many lines of sight, and those for all the clusters
together.  
\subsection{Trends for all clusters considered together}
We take the union of the estimated masses of all clusters and all lines of sight
(aside from those discarded as discussed in \S 2) and repeat the
analysis
of \S 3.   Now $\vec{M}_{\rm
  obs}/\mtrue= \vec{M}_{\rm est}/\mtrue - \langle \vec{M}_{\rm
  est}/\mtrue \rangle$.
The average $\langle \vec{M}_{\rm est}/\mtrue \rangle$ is 
over all clusters and lines of sight, and for our case is within a few
percent of 1. 
The resulting covariance matrix for these mass scatters is shown in Table \ref{tab:fullcov}.
The covariances for the combined
sample are larger than the averages for individual clusters, the latter are
shown in Table \ref{tab:fullcov} in parentheses for comparison, and were
plotted
individually in Fig.~\ref{fig:covscatter}.
The increased scatter is not unexpected as it is a combination of
scatters from objects which are all themselves scattered around
different
central values.
\begin{table*} 
   \centering
\begin{tabular}{cccccc} 
& $\mnred/\mtrue$ & $\mnph/\mtrue$ & $\msz/\mtrue$ & $\mvel/\mtrue$ & $\mwl/\mtrue$ \\
$\mnred/\mtrue$ & 0.18 (0.08) & 0.06 (0.02) & 0.05 (0.04) & 0.04 (0.03) & 0.04 (0.03) \\ 
$\mnph/\mtrue$ & 0.06 (0.02) & 0.04 (0.02) & 0.02 (0.01) & 0.02 (0.01) & 0.02 (0.01) \\ 
$\msz/\mtrue$ & 0.05 (0.04) & 0.02 (0.01) & 0.09 (0.07) & 0.04 (0.02) & 0.04 (0.01) \\ 
$\mvel/\mtrue$ & 0.04 (0.03) & 0.02 (0.01) & 0.04 (0.02) & 0.24 (0.20) & 0.05 (0.02) \\ 
$\mwl/\mtrue$ & 0.04 (0.03) & 0.02 (0.01) & 0.04 (0.01) & 0.05 (0.02) & 0.11 (0.05) \\ 
\end{tabular}
   \caption{Covariance matrix for full set of measurements $\vec{M}_{\rm
       obs}/M_{\rm true}$, for all clusters, and, in parentheses, the average of the individual
     cluster covariances in
     Fig.~\ref{fig:corrscatter}.  The median values for the the
     individual cluster covariances tend
     to be smaller than the average values. 
The covariances for the full sample tend to be larger than the
averages cluster to cluster.
}
\label{tab:fullcov}
\end{table*}
The corresponding PC
eigenvalues are $\lambda_{i,M,total}$=(0.31,  0.18,  0.094,  0.061,
0.022). 
The direction of largest scatter has $\lambda_{0,M,total}/\sum{\lambda}
\sim$0.47, compared to the average of 0.66
for individual clusters
shown in
Fig.~\ref{fig:pccontrib}.  The total mass scatter variance $\sum{\lambda} = 0.67$, which is relatively
high compared to 
 the average for individual clusters; the product $\prod \lambda$ is also
 relatively larger
(7$\times 10^{-6}$), as to be expected from the increased covariances
in Table \ref{tab:fullcov}. \footnote{
For the 70 clusters with mass $\geq 2\times 10^{14} h^{-1} M_\odot$, the total
scatter $\sum{\lambda}$ goes down (to 0.44 from 0.67 for the sample with $M \geq
10^{14} h^{-1} M_\odot$), and the direction of the
$\hat{PC}_{0,M,total}$ slightly rotates as seen in Table
\ref{tab:pcvecs}.  %
The fraction of variance in $\lambda_{0,M}$
increases, i.e.  the direction of the largest
scatter has more of the scatter.}
The directions of largest scatters are similar to those for the clusters
considered separately (Table \ref{tab:pcvecs}), although
$\hat{PC}_{4,M,total}$ has a much smaller SZ component
than $\hat{PC}_{4,M,minsq}$.
The correlations with $\hat{PC}_{0,M,total}$ of the projections of the
different mass
observables ($\mnred$, $\mnph$, $\msz$, $\mvel$, $\mwl$)
are (0.69, 0.57,  0.57, 0.79, 0.57), similar to the median of
their individual cluster counterparts,
 (0.69, 0.51, 0.48, 0.87,  0.47), shown in
Fig.~\ref{fig:corrmpc}.
The correlations are fairly large for 
the other $\hat{PC}_i$ components, but these by definition contribute less to the total mass scatter.
$\hat{PC}_{0,M,total}$ and $\Delta \vec{M}$
are $23^\circ$ 
apart, similar to their counterparts for the individual clusters.
Again, just as for individual clusters,
the closer one is to the long axis of the cluster,
the larger the fraction of scatter due to
$\hat{PC}_{0,M,total}$ ($\sim$0.4 correlated, see,
e.g. Fig.~\ref{fig:vecprop} for the individual cluster distribution).  
The next leading correlations of $\hat{PC}_{0,M,total}$ are with the position of 
the largest subgroup and the direction perpendicular to the cluster's
dominant filamentary plane or mass plane.   These trends were very
close to those found for clusters individually.  These relations are
still however based on the estimated mass as a function of true mass,
and thus not
immediately applicable to observational samples, as we now discuss.

\subsection{Future extensions to observational samples}
One can also consider using some variant of our PCA analysis for an
observational sample, which would have some range of true
cluster masses and some observations, as we do in our box.  However,
our implementation of PCA doesn't directly carry over, and our
measurement sample in hand is not appropriate.  We
discuss these limitations and possible ways forward here.

Observations are concerned with $\mtrue$ as a function of
estimated mass, $\mtrue(M_{\rm est})$.  An observer does not have
$\mtrue$.  We have instead
been calculating $M_{\rm est}(\mtrue)$ and in addition dividing by $\mtrue$. 
Applying PCA in a way useful to observations (so that one can take 5
methods to measure mass of one cluster and compare to the PC vectors
calculated in simulation)
requires PC vectors calculated, from simulation, for a representative range of $M_{\rm
  est}$ and $\mtrue$, and a proxy for the unobservable
$\mtrue$.\footnote{Using no proxy for
  $\mtrue$, i.e. doing PCA on $M_{\rm est}$ alone,
differently weights the scatter of high and low
mass clusters.  In our sample, caveats below, most of the scatter
then corresponds to all mass estimates increasing or decreasing together
in equal amounts.  Projections on this combination of mass scatters are weakly correlated
with observations along the long axis of the cluster, and less
correlated with other directions.
}
As a proxy for $\mtrue$, one possibility is the likelihood mass.  This
would come
from simulations, which are already required to calibrate covariances and offsets.
Another possible mass proxy uses the principal component vectors
directly, in principle encoding similar information.
That is, one has from the definition of PC vectors,
\begin{equation}
  \frac{\vec{M}_{\rm obs}^\alpha}{\mtrue} =\frac{\vec{M}_{\rm
      est}^\alpha}{M_{\rm true}}-\left \langle\frac{\vec{M}_{\rm est}}{M_{\rm true}}\right\rangle   = \sum_i
  a^\alpha_i \hat{PC}_{i,M} \; .
\end{equation}
The quantities aside from $\mtrue$ are either measured , e.g.,
$\vec{M}_{\rm est}^\alpha$, or calculated from simulations, e.g.,
$\langle \vec{M}_{\rm est}/\mtrue \rangle, \hat{PC}_{i,M}$.  In
particular, we have
\begin{equation}
\hat{PC}_{4,M,total} \cdot\left(\frac{\vec{M}_{\rm est}^\alpha}{M_{\rm
    true}}-\left\langle \frac{
  \vec{M}_{\rm est}}{M_{\rm true}}\right\rangle\right)  = 
a_4^\alpha 
\end{equation}
but the variance of $a_4$ is given by $\lambda_{4,M,total}$.  If $\lambda_{4,M,total}$ is
very small then we can try the approximation $a_4 \sim
0$ , so that projecting on $\hat{PC}_{4,M,total}$ gives 
the approximation\footnote{An extreme and unphysical limit of this case would be
  if one measurement method had tiny scatter and no correlation with the
  other measurement methods.  In this case $\hat{PC}_{4,M,total}$ would be
  proportional to a measurement via this one method, and the last equation would
  basically give that the estimated mass in this method is the true
  mass; up to any overall biases that might exist, an unbiased mass
  estimator would make the denominator 1.}
\begin{equation}
M_4 \sim \frac{ \hat{PC}_{4,M,total} \cdot\vec{M}^\alpha_{\rm est}} {\hat{PC}_{4,M,total} \cdot \langle\frac{
  \vec{M}_{\rm est}}{M_{\rm true}}\rangle}\; .
\end{equation}
The general use of this
approach
depends upon the size of $\lambda_{4,M,total}$ for the system; the
smaller $\lambda_{4,M,total}$ is, 
the better it appears this approximation should work.  As there
tend to be some minus signs in $\hat{PC}_4$,
some catastrophic failures will occur where
$M_4 \ll M_{\rm est,i}$, for all $i$.

However, the relevance of testing these possible $\mtrue$ proxies on our sample
seems limited.
In particular, the $P(M_{\rm est}|\mtrue)$ for our sample (all $M \geq
10^{14} h^{-1}M_\odot$ halos observed along $\sim 96$ lines of sight
each) is not representative of any expected observed sample, nor is
our $P(\mtrue)$, which is required for likelihoods.
Observing our few high mass clusters along many lines of sight is not
a good approximation to observing many clusters along one
line of sight:  the few high mass clusters observed along many lines of
sight in particular do not well sample the realistic population of
high mass clusters. There are no clusters appearing fewer than
$\sim 96$ times; in particular, rare high mass clusters which would be expected
from the number of lower mass clusters ``present'' are missing.
Halos with the same $M_{\rm est}$ (the only observable) but a lower
$\mtrue$
will also occur in an observational sample, and might contribute differently to the scatter as well.
It would also be important to include the neglected,
in our simulation, line of sight larger scale
scatter for SZ and weak lensing and the systematics mentioned above.
These would be very interesting directions to pursue in future work.

In summary, in this section we considered all the clusters in the box, along all
lines of sight, to see how cluster-to-cluster variation altered mass
scatter relations found in earlier sections for individual clusters; 
the trends remained but changed
in strength.  In
particular, the dominant mass scatter combination, similar in form to
that for many of the individual clusters, still seems to 
be more prevalent when
looking down the long axis of the cluster or perpendicular to the mass
or filament plane of the cluster.  
In the second subsection,
we mentioned two possible methods for extending our
analysis which do not require prior knowledge of $\mtrue$, replacing
it with the likelihood mass or a mass derived from
$\hat{PC}_4$.   These would be interesting to apply to an
observational sample, but would need a very closely matched simulation.
It would be interesting to see if the resulting PC vectors and
projections on them by estimated mass measurement methods have correlations
with cluster orientation or perhaps $M_{\rm proxy}/\mtrue$.

\section{Outliers}
We have focussed on general trends above, but not all clusters (or all
lines of sight) fell on the general trends.  We searched for
properties of outliers or tails in the distributions of the various
quantities related to $\lambda_{i,M}$, outliers in average cluster measured
mass vs. true cluster mass, clusters with different maximum or
minimum covariance mass measurement method pairs than the majority of
clusters, and clusters where
$\hat{PC}_{0,M}$ had at least one opposite sign coefficient with absolute
value $\geq 0.1$ (i.e. largest direction of scatter not corresponding
to all mass scatters increasing together, mentioned earlier).

The various outliers did not seem to follow
any pattern.  Some outliers were common to a cluster, e.g. sometimes a large $\lambda_{4,M}$
outlier occurred with a large $\prod \lambda$, as might be expected, as
$\lambda_{4,M}$ is the smallest of the $\lambda_{i,M}$.
Some of the clusters which had $\hat{PC}_{0,M}$ not aligned with the
most likely $\hat{PC}_{0,M,minsq}$ had much larger
contributions to
mass scatter from Compton decrement
than the average cluster, often due to close massive halos,
and
$\lambda_{0,M}/\sum{\lambda}$ tending to be smaller than for the
usual cluster.  
The clusters having different correlations of lines-of-sight
properties with mass measurement methods don't seem to have a clear relation to the outliers in mass scatter properties.

\section{Summary and discussion}

The scatters between estimated and true cluster masses, for different
observational methods, are often
correlated.  Understanding these correlations
is becoming more and more important as reliance on multiwavelength measurements
increases.  For instance,
correlations and covariances in scatters affect both error estimates
in multiple measurements of individual clusters and produce a bias in
measurements
of stacked objects (see \citet{Ryk08,Sta10,WCS,Ang12} for
detailed discussion).  (Using the full
covariance matrices has been done in some cluster analyses, e.g. by
\citet{Roz09,Man10,Ben11}.)
We characterized the scatter and
correlations of clusters in two ways.

We started by considering clusters individually to identify mass scatter
properties due to line of sight effects for 
243 clusters in an N-body simulation, each along $\sim 96$ lines of sight.
We used five observational mass proxies:  red galaxy richness, phase space
galaxy richness, Sunyaev-Zel'dovich flux, velocity dispersions and
weak lensing.  It would also be very interesting to include
  X-ray observation as well, but our attempt at a proxy, based on
  assigning fractional X-ray flux to cluster galaxy subgroups, was not
  illuminating.  These are employed to find cluster masses
in current and upcoming large scale cluster surveys.

For each cluster, we characterized the ``shape''  and ``volume'' of the mass scatters
of $M_{\rm est}$ calculated as a function of $\mtrue$,
using PCA, or Principal Component Analysis, to obtain
a set of non-covariant measurements.
Most clusters had one combination of observational mass scatters contributing the
majority of the mass scatter, and this combination was similar for
many of the clusters, i.e. they had a similar largest principal
component $\hat{PC}_{0,M}$.
This scatter combination was a larger fraction of the total line of sight mass
scatter when the cluster was 
observed along the long axis of the cluster.  Weaker relations with observations along other
cluster intrinsic and environmental axes were seen. Identifying the
long axis of course requires the clusters
to have non-spherical shapes.  In our case the cluster member dark
matter particles
were determined using the FoF finder with linking length
$b=0.168$.  

Individual cluster mass scatter properties due to line of sight effects were
then compared to several intrinsic and environmental cluster
properties, including triaxiality, planarity of
filament or halo mass in the immediate neighborhood of the cluster and
relative richness of largest galaxy subgroup within the cluster.
For example, pairwise correlations, and their combined effects using
PCA, showed
that clusters with average mass measurements (over lines of sight)
which are large tend to also have large mass scatter around their
average, relatively high sphericity and small triaxiality, richness in
the largest subgroup, and offset of the galaxy position average from
the cluster center.
 Relations were also seen
for other quantities such as
concentration, recent major merger time and
fraction of halo mass near the cluster within a 3 $h^{-1} Mpc$ plane.

Finally, instead of considering each cluster individually, we
considered the sample of all clusters and all lines of sight
together, and found that most of the trends for the analysis of
$\vec{M}_{\rm est}/\mtrue$ remained, albeit at
different strengths.   The projection on the direction of largest
scatter was more weakly correlated with the observation
angle relative to the cluster long axis.

It is interesting to think about applying these methods directly to
observation.  One would need more information, in particular
estimates of $\mtrue(M_{\rm est})$, rather than the opposite which we
have here.   This
requires calibrations from simulations which better
sample an expected observational sample at the high mass end, and
which also include estimated masses from lower mass halos (as well as
faithfully reproducing the observational systematics and selection
function).   Such a simulation would provide correlations and
covariances and predicted
likelihood masses $M_{\rm like}$ (and a variant, $M_4$ considered
above, based on a narrow direction of scatter in our sample).  PCA
could then be applied to 
$M_{\rm est}/M_{\rm like}$ or $M_{\rm est}/M_4$, rather than $M_{\rm
  est}/M_{\rm true}$ as we did here.  It would be interesting to see
if these PC vectors also had relations to cluster orientation such as
we found, or
perhaps other quantities such as $M_{\rm like}/\mtrue$.  They might
also give some idea
of which follow up mass measurement methods would together provide the
most
constraining power.  One could for example identify
the measurement method with potentially the least covariance with
measurements
already in hand.  More generally one could design a combination of
measurement 
methods with smaller
covariant
scatter (and hopefully smaller scatter as well) using calculated PC vectors as a guide, if the simulations
are faithful enough.  It is a major challenge to
accurately capture the systematics and selection
function of observational surveys with numerical simulations.
Another technical issue is to improve estimates
of the correlations and covariances, so that sets of inconsistent
measurements can be more easily recognized. 
It would be very interesting to do such analyses on a larger box, and/or
with other measurement methods.

\section*{Acknowledgements}

YN thanks E. Rasia for discussions 
and the Essential Cosmology
for the Next Generation school for the opportunity to present results
of this work.  She was supported in part by NSF.
JDC thanks A. Ross for suggestions and
M. White for numerous helpful discussions and suggestions.
We both thank Z. Lukic, E. Rozo and the anonymous referee for
extremely helpful criticisms and comments on
the draft.

\end{document}